\documentclass[journal,twocolumn,10pt]{IEEEtran}
\normalsize
\usepackage[noadjust]{cite}
\usepackage{filecontents}
\usepackage{color,colortbl}
\usepackage[pdftex]{graphicx}
\usepackage{verbatim} 
\usepackage{epstopdf}
\usepackage[cmex10]{amsmath}
\usepackage[tight,footnotesize]{subfigure}
\usepackage{pifont}
\usepackage{multirow}
\usepackage{multicol}
\usepackage{epstopdf}
\usepackage{graphicx}
\usepackage{xcolor}
\usepackage{tikz}

\usepackage[colorlinks=true,pdfstartview=FitV,linkcolor=black,citecolor=black, urlcolor=blue,plainpages=false]{hyperref}
\usepackage{hyperref}
\usepackage{amssymb}
\definecolor{Gray}{gray}{0.99}
\newcommand{\E}{{\rm I\kern-.3em E}}
\usepackage{lipsum}
\usepackage{mathtools}
\usepackage{cuted}
\usepackage{soul}
\usepackage{makecell}

\DeclarePairedDelimiter\ceil{\lceil}{\rceil}
\newcommand*{\taghere}[1][0pt]
{\ifmeasuring@\else
  \global\tag@heretrue
  \tikz[remember picture,overlay]{\coordinate (taghere) at (0pt,#1);}%
\fi}

\def\place@tag{%
    \iftagsleft@
      \kern-\tagshift@
      \iftag@here
        \global\tag@herefalse
        \tikz[remember picture,overlay]%
          {\path (taghere) -| node[anchor=base]{\rlap{\boxz@}} (0pt,0pt);}%
      \else
        \if1\shift@tag\row@\relax
            \rlap{\vbox{%
                \normalbaselines
                \boxz@
                \vbox to\lineht@{}%
                \raise@tag
            }}%
        \else
            \rlap{\boxz@}%
        \fi
        \kern\displaywidth@
      \fi
    \else
      \kern-\tagshift@
      \iftag@here
        \global\tag@herefalse
        \tikz[remember picture,overlay]%
          {\path  (taghere) -|  node[anchor=base]{\llap{\boxz@}} (0pt,0pt);}%
      \else
        \if1\shift@tag\row@\relax
            \llap{\vtop{%
                \raise@tag
                \normalbaselines
                \setbox\@ne\null
                \dp\@ne\lineht@
                \box\@ne
                \boxz@
            }}%
        \else \llap{\boxz@}%
        \fi
      \fi
    \fi
}
\makeatother

\usepackage{etoolbox}
\makeatletter 
\pretocmd\@bibitem{\color{black}\csname keycolor#1\endcsname}{}{\fail}
\newcommand\citecolor[1]{\@namedef{keycolor#1}{\color{black}}}
\makeatother

\begin{document}
\title{Digital Predistortion for \\ Multiuser Hybrid MIMO at mmWaves}  
\author{Alberto~Brihuega,~\IEEEmembership{Student Member,~IEEE,}
        Lauri~Anttila,~\IEEEmembership{Member,~IEEE,}
        Mahmoud~Abdelaziz,~\IEEEmembership{Member,~IEEE,}
        Thomas~Eriksson,~\IEEEmembership{Member,~IEEE,}
        Fredrik~Tufvesson,~\IEEEmembership{Fellow,~IEEE,}
        and Mikko~Valkama,~\IEEEmembership{Senior~Member,~IEEE}
\thanks{This work was financially supported by the Academy of Finland (grants {304147, 301820, 326405 and 319994}) and Tampere University Doctoral School.}
\thanks{A.~Brihuega, L.~Anttila, and M.~Valkama are with the Department of Electrical Engineering, Tampere University, Tampere, Finland.} 
\thanks{M.~Abdelaziz is with Zewail City of Science and Technology, Egypt.}
\thanks{T.~Eriksson is with the Department of Electrical Engineering, Chalmers University of Technology, Gothenburg, Sweden.}
\thanks{F.~Tufvesson is with the Department of Electrical and Information Technology, Lund University, Lund, Sweden.}
\thanks{{This paper has supplementary downloadable material available at http://ieeexplore.ieee.org., provided by the authors. The material includes measured 28~GHz PA models for two different types of state-of-the-art power amplifiers. This material is 95~kB in size.}}
}
\maketitle
\begin{abstract}
Efficient mitigation of power amplifier (PA) nonlinear distortion in multi-user hybrid precoding based broadband mmWave systems is an open research problem. In this article, we carry out detailed signal and distortion modeling in broadband multi-user hybrid MIMO systems, with a bank of nonlinear PAs in each subarray, while also take the inevitable crosstalk between the antenna/PA branches into account. Building on the derived models, we adopt and describe an efficient closed-loop (CL) digital predistortion (DPD) solution that utilizes only a single-input DPD unit per transmit chain or subarray, despite crosstalk, providing thus substantial complexity-benefit compared to the state-of-the art multi-dimensional DPD solutions. We show that under spatially correlated multipath propagation, each single-input DPD unit can provide linearization towards every intended user, or more generally, towards all spatial directions where coherent propagation is taking place, {and that the adopted CL DPD system is robust against crosstalk}. Extensive numerical results building on practical measurement-based mmWave PA models are provided, demonstrating and verifying the excellent linearization performance of the overall DPD system in different evaluation scenarios.
\end{abstract}


\IEEEpeerreviewmaketitle

\section{Introduction}
\IEEEPARstart{T}{he} demands for higher data rates and larger network capacities have led mobile communications system evolution to adopt new spectrum at different frequency bands, to deploy larger and larger antenna arrays, and to substantially densify the networks \cite{Intro_1, Intro_0, Intro_5G,MIMO_Intro_SP,ScalingMIMO}. Millimeter wave (mmWave) communications allow to leverage the large amounts of available spectrum in order to provide orders of magnitude higher data rates, but also impose multiple challenges compared to sub-6 GHz systems. In general, the propagation losses at mmWaves are considerably higher than those at sub-6 GHz bands, and thus large antenna gains are typically needed at both the transmitter (TX) and receiver (RX) ends in order to facilitate reasonable link budgets \cite{Intro_1, Intro_0, Intro_5G,Intro_HybridMIMO}. 

Operating at mmWave frequencies allows to pack a large number of antennas in a small area. However, the implementation of fully digital beamforming based large antenna array transmitters turns out to be very costly and power consuming \cite{Intro_3}. For this reason, many works have proposed and considered hybrid analog-digital beamforming solutions \cite{HybridPrecoding1,HybridPrecoding2,Hybrid_precoding_new1,Intro_3,Intro_HybridMIMO,Intro_HybridMIMO2,Precoding1,AE1,AE2}  as a feasible technical approach and compromise between implementation costs, power consumption, and beamforming flexibility. This is also well in-line with the angular domain sparsity of the mmWave propagation channels \cite{Intro_5G,Precoding1, mmWaveCH, ChannelCorr,AE2}, which results in reduced multiplexing gain. In general, there are several hybrid architectures depending on how the analog beamforming stage is implemented \cite{Intro_HybridMIMO,Intro_3}. Two common architectures are the so-called full-complexity architecture, where an individual analog precoder output is a linear combination of all the RF signals, and the so-called reduced-complexity architecture, in which each TX chain is connected only to a subset of antennas, known as a subarray. The reduced complexity architecture is known to be more feasible for practical implementations \cite{Intro_3,Intro_HybridMIMO,Intro_HybridMIMO2,Reduced_HybridMIMO} and is thus assumed also in this article.

\vspace{-3mm}
\subsection{Nonlinear Distortion and State-of-the-Art}
In general, energy efficiency is an important design criterion for any modern radio system, including 5G and beyond cellular
systems \cite{Intro_1}. Therefore, in the large array transmitter context, efficient operation of the power amplifier (PA) units is of key importance. To this end, highly nonlinear PAs operating close to saturation are expected to be used in the base stations (BS) \cite{GreenComm}, {leading to nonlinear distortion that degrades the received signal quality of both the in-channel intended receivers as well as the neighboring channel victim receivers}. Nonlinear distortion due to PAs in massive MIMO type transmitter systems has been  studied in the recent literature \cite{OOB_Mollen,OOB_Emissions_Ours,Prediction_distortion,Intro_7}. Specifically, in \cite{OOB_Mollen}, it was shown that the worst case emissions occur in the direction of the intended receiver,
since out-of-band (OOB) distortion also gets beamformed towards this direction. 

Compared to simply backing off the PA input power, a much more efficient approach to control the PA-induced emissions while still operating close to saturation is to utilize digital predistortion (DPD) \cite{Trad_DPD,Trad_DPD2}. DPD has been recently studied in the context of large antenna arrays and MIMO systems in, e.g., \cite{DPD_MM_1,DPD_DigitalMIMO,DPD_MM_2,DPD_MM_3,DPD_MM_4,DPD_MM_5,DPD_MM_6,OTA_combining_DPD,OTA_combining_DPD2,8605680,MIMO_DPD_2,MIMO_DPD_3,MIMO_DPD_4,Crosstalk_Chalmers}. In \cite{DPD_MM_1, DPD_DigitalMIMO,MIMO_DPD_2,MIMO_DPD_3,MIMO_DPD_4}, fully digital beamforming based system was investigated. In \cite{DPD_MM_1}, a dedicated DPD unit per antenna/PA was considered, primarily focusing on the reduction of the complexity of the DPD learning algorithm. However, a dedicated DPD unit per antenna/PA branch may not be implementation-feasible in large array transmitters because of the complexity and power consumption issues. Therefore, in \cite{DPD_DigitalMIMO} the authors proposed an alternative DPD solution where a single DPD unit can linearize an arbitrarily large antenna array, with multiple PAs, when single-user phase-only digital precoding is considered. {The MIMO DPD works \cite{MIMO_DPD_2,MIMO_DPD_3,MIMO_DPD_4}, in turn, are specifically seeking to tackle the antenna crosstalk problem, which can be considerable especially with closely-spaced antenna elements in large array transmitters \cite{Swedes_review,28GHz_crosstalk}. To this end, different indirect learning architecture (ILA) based multi-dimensional DPD solutions have been proposed. However, the amount of the involved basis functions (BFs) and thus the complexity of such multi-input DPD solutions commonly grows steeply with the number of TX paths, being thus prohibitive from the implementation point of view in practical systems, particularly with more than two TX chains.}

In \cite{DPD_MM_2,DPD_MM_4,DPD_MM_5,DPD_MM_6,OTA_combining_DPD,OTA_combining_DPD2}, DPD solutions for single-user hybrid MIMO transmitters were investigated assuming the reduced-complexity architecture \cite{Intro_3}. To this end, and since each DPD unit operates in the digital domain, an individual predistorter is responsible for linearizing all the PAs within its respective subarray. Since the PA units are in practice mutually different, this is essentially an under-determined problem and generally yields reduced linearization performance, when compared to linearizing each PA individually. In \cite{DPD_MM_2}, the DPD learning is based on observing only a single PA output, within each subarray, while the works in \cite{DPD_MM_3,8605680} consider the multiuser case but adopt a simplifying assumption that all the PAs are mutually identical. As a result, both approaches lead to reduced linearization performance, due to the mutual differences between real PA units and their exact nonlinear distortion characteristics. Additionally, only a third-order PA model and corresponding DPD processing are considered in \cite{DPD_MM_3}.  

Other works on hybrid MIMO \cite{DPD_MM_5,DPD_MM_6,OTA_combining_DPD,OTA_combining_DPD2,Low_number_steering,OTA_feedback_DPD} seek to benefit from the spatial characteristics of the OOB emissions in array transmitters in order to develop efficient DPD solutions. These works rely on the fact that unwanted emissions are more significant in the direction of the intended receiver. In the single-user case, the received signal of the intended user under line-of-sight (LOS) propagation can be mimicked by coherently combining all the individual PA output signals within the subarray. This forms the signal for DPD parameter learning and overall effectively yields a well defined single-input-single-output DPD problem. Such DPD processing results in minimizing the OOB emissions in the direction of the intended receiver \cite{DPD_MM_5}. {The works in \cite{DPD_DigitalMIMO,DPD_MM_2,DPD_MM_3,DPD_MM_4,DPD_MM_5,DPD_MM_6,OTA_combining_DPD,OTA_combining_DPD2, 8605680,Low_number_steering,OTA_feedback_DPD} either assume single-user transmission or adopt some other simplifying assumptions such as all PAs being identical, pure LOS propagation or narrowband fading. Thus, nonlinear distortion modeling and corresponding DPD solutions for true multi-user hybrid MIMO systems under mutually different PA units and broadband channels have not been pursued in the current literature. Moreover, none of the works \cite{DPD_MM_1,DPD_DigitalMIMO,DPD_MM_2,DPD_MM_3,DPD_MM_4,DPD_MM_5,DPD_MM_6,OTA_combining_DPD,OTA_combining_DPD2,8605680} tackle the problem of crosstalk, which, on top of the additional implementation related challenges described in \cite{MIMO_DPD_2,MIMO_DPD_3,MIMO_DPD_4}, also causes an effective loading phenomenon such that the behavior of the individual PAs are dependent on the array steering angle \cite{Swedes_review, Low_number_steering}. This is known as load modulation, and may necessitate continuous learning of the DPD filter coefficients in order to keep track of the fast beam changes \cite{Fager17}. These are the open challenges that this article is addressing.}

\subsection{Contributions and Novelty}
In this article, we first provide detailed signal and distortion modeling for hybrid-precoded multi-user MIMO-OFDM systems under nonlinear and mutually different PAs {affected by input and output crosstalk} and containing memory. For generality, the signal models cover both the classical single-beam as well as the more elaborate multi-beam \cite{AnalogBeamforming} \cite{Heath_multibeam} analog beamforming, per subarray. Building on the derived signal and distortion models, we then adopt a closed-loop (CL) single-input DPD solution, structurally similar to that considered in \cite{DPD_MM_5} in single phased-array and pure LOS propagation context, while being here deployed in parallel in each TX chain. Due to the considered hybrid precoding based multi-user transmission scenario, the received signals by the intended and potential victim users are contributed by the transmitted signals from all the subarrays, particularly in the multi-beam case but also in the single-beam case due to finite spatial isolation. As a consequence, the overall DPD system needs to provide linearization not only to a single point in space, as was the case in \cite{DPD_MM_5,DPD_MM_6}, but to multiple points and corresponding receivers -- something that is shown to be feasible in this article through the analytical signal models.  

For parameter estimation purposes, the PA output signals, per each subarray, are coherently combined in the RF domain, similar to \cite{DPD_MM_5,DPD_MM_4},
reflecting the radiated nonlinear distortion of each subarray. Importantly, it is analytically shown that such an approach can provide sufficient feedback knowledge for parameter estimation purposes, despite the broadband propagation impacting the true received signals.
Additionally, in this article, the inherent robustness of the CL system against antenna crosstalk is established, and leveraged to linearize the transmitter system by means of single-input DPD units. Owing to the CL operating principle and computationally simple learning algorithms, the considered methods may allow for even continuous parameter learning and tracking, which can be essential in mmWave systems with fast beam changes and corresponding changes in the radiated nonlinear distortion \cite{Swedes_review}, \cite{Fager17}. Additionally, compared to the existing multi-dimensional DPD methods and associated extended sets of BFs \cite{MIMO_DPD_2,MIMO_DPD_3,MIMO_DPD_4}, the considered single-input DPD units, applied per TX chain or subarray in this work, provide largely reduced complexity and thus more implementation feasible solution in terms of hardware and energy consumption. 

Overall, the scope, contributions and novelty of this article can be summarized as follows:
\begin{itemize}
    \item instead of a single phased-array, linearization of true hybrid MIMO transmitter with multiple simultaneously operating subarrays with digital precoding and analog beamforming is pursued;
    \item nonlinear distortion related received signal and system models are derived, incorporating mutually different wideband PA units, crosstalk effects and broadband multipath fading, while also considering both the single-beam and multi-beam analog beamforming cases for generality;
    \item it is analytically shown that by adopting the single-input closed-loop DPD units, developed in \cite{DPD_MM_5} in the single phased-array and LOS propagation context, in all parallel transmitters, nonlinear distortion can be efficiently mitigated despite the strong PA input and PA output crosstalk;
    \item linearization of a strongly nonlinear multi-user mmWave hybrid array system is successfully shown, through comprehensive numerical examples, containing both single-beam and multi-beam analog beamforming cases, wideband measurement based mmWave PA models, crosstalk and imperfect channel knowledge;
    \item the results also show that true over-the-air (OTA) feedback is not needed, as the hardware-based subarray-level combiners can provide sufficient feedback knowledge for successful linearization without incorporating the true channel responses.
\end{itemize}
To the best of the authors' knowledge, this is the first article that successfully shows and demonstrates DPD-based linearization of an mmWave multi-user hybrid beamforming based array system with a bank of nonlinear PA units and broadband propagation.

The remainder of this paper is organized as follows: In Section \ref{System_Model}, the hybrid multiuser MIMO linear system model considered in this work is described. In Section \ref{Nonlinear_model}, the modeling and analysis of the nonlinear distortion arising from the nonlinear PAs are carried out, with specific emphasis on the combined or observable distortion. Then, Section \ref{DPD structure} describes the single-input DPD structure and parameter learning solution, {while the extension to the challenging crosstalk case is provided in Section \ref{sec:crosstalk}}. {Section \ref{sec:complexity} provides then a complexity analysis and assessment of the considered CL-DPD and the reference ILA-based DPD}. In Section \ref{Results}, the numerical performance evaluation results are presented and comprehensively analyzed. Lastly, Section \ref{sec:Conclusions} will provide the main concluding remarks, {while selected details are provided in the Appendix.}

\section{Multiuser Hybrid MIMO - Linear System Model}\label{System_Model}
\subsection{Basics}
\begin{figure*}[t!]
\centering
\includegraphics[width=.85\linewidth]{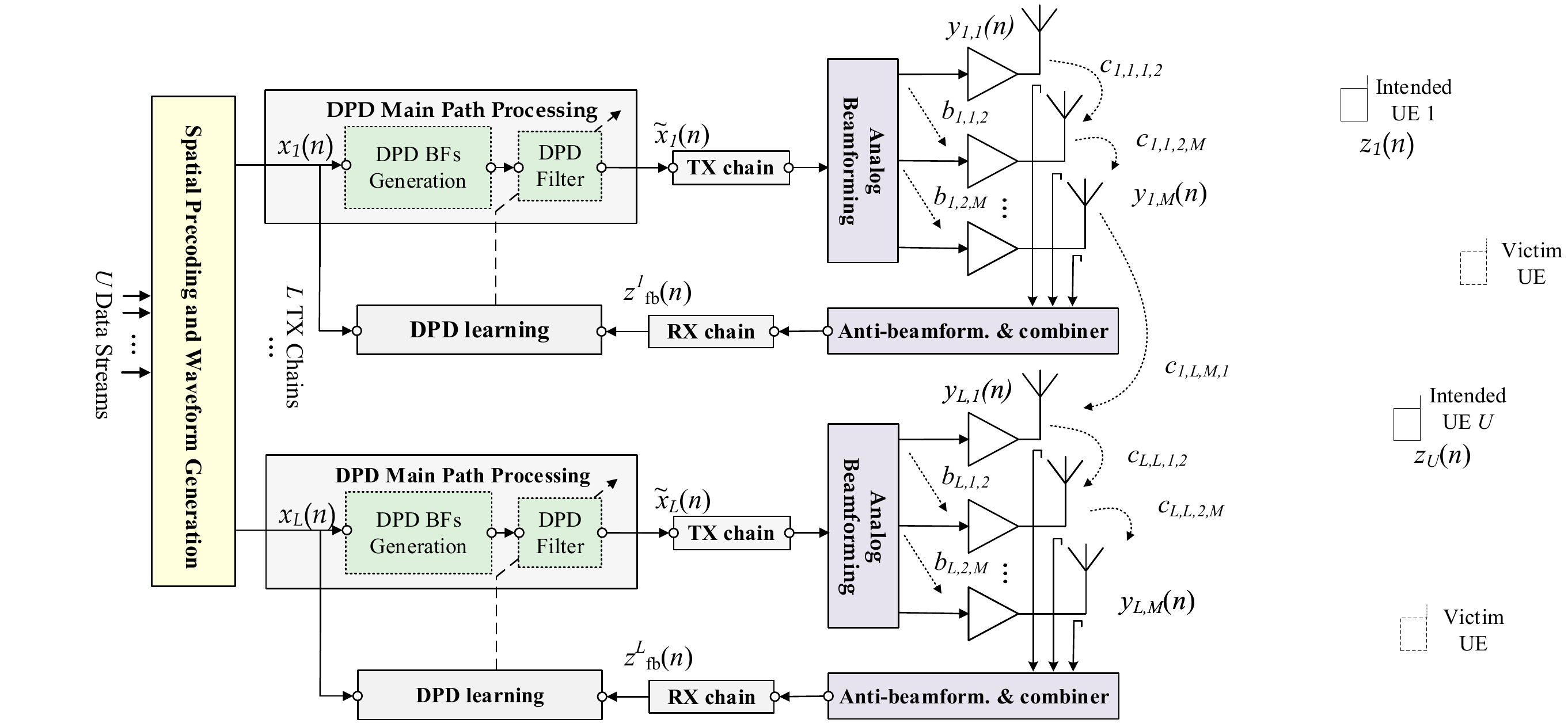}
\caption[]{Block diagram of the considered hybrid beamforming based multiuser MIMO-OFDM transmitter, {illustrating also different crosstalk mechanisms at the inputs and outputs of the power amplifiers. Signals ${x}_l(n)$ and $\tilde{x}_l(n)$ are the input and output of the DPD system in subarray $l$. Overall, $U$ intended UEs are served, through spatial multiplexing, while also some victim UEs operating in the adjacent channels are shown.}}
\label{fig:WholeSystem}
\end{figure*}

The overall considered hybrid beamforming based multiuser MIMO-OFDM transmitter is shown in Fig. \ref{fig:WholeSystem}, containing $L$ TX chains and $M$ antenna units per subarray, while serving $U$ single-antenna users simultaneously. The subcarrier-wise BB precoder, denoted as $\mathbf{F}[k]$ $\in \mathbb{C}^{L \times U}$ at subcarrier $k$, is responsible for mapping the $U$ data streams onto $L$ TX chains and for spatially multiplexing the different users, while the RF precoder, $\mathbf{W}$ $\in \mathbb{C}^{{LM} \times L}$, focuses the energy towards the dominant directions of the channel. It is further assumed that $U \leq L \leq LM$. The precoded data symbol blocks are then transformed to time-domain OFDM waveforms through IFFTs of size $K_{\mathrm{FFT}} > K_{\mathrm{ACT}}$ where $K_{\mathrm{ACT}}$ denotes the number of active subcarriers.
 A cyclic prefix is then added to the sample blocks.
 {Finally, it is noted that RX side beamforming can in practice be adopted to improve the received signal strength. However, as the linear signal and the nonlinear distortion exhibit the same beamforming gain, there are no major differences compared to the single-antenna RX case when it comes to assessing the TX-induced nonlinear distortion behavior. Consequently, no RX beamforming is considered in this work for notational simplicity.}

\subsection{mmWave Channel Model}
{In order to accurately incorporate the frequency-selectivity as well as the spatial correlation characteristics of the array channels, we adopt an extended version of the geometric Saleh-Valenzuela channel model \cite{SalehValenzuela}, similar to \cite{HybridPrecoding1}}.  
Specifically, we assume a clustered channel model with $C$ clusters, where each cluster is made up of $R$ rays. Each cluster $c$ has a certain path-delay $\tau_{c}$,  angle of arrival {$\phi_{c}$}, and angle of departure {$\theta_{c}$}, while each ray has its corresponding ray-delay, angle of arrival and angle of departure denoted by $\tau_r$, {$\upsilon_r$} and {$\nu_r$}, respectively. Lastly, let {$f_{\mathrm{flt}}(t)$} denote a band-limitation function. Following the above mentioned model, the $T_s$-spaced delay-$d$ channel vector \cite{HybridPrecoding1} for the $u$-th user reads then 
\begin{equation}
    \mathbf{h}_{u}[d] = \sum_{\substack{c=1}}^{C} \sum_{\substack{r=1}}^{R} {\rho_{c,r}}f_{\mathrm{flt}}(dT_{\mathrm{s}} -\tau_{c} - \tau_{r}){a}_{\mathrm{Rx}}(\phi_{c}-\upsilon_r)\mathbf{a}_{\mathrm{Tx}}(\theta_c - \nu_r) \label{channel_model}
\end{equation}
where {$\rho_{c,r}$} is the complex gain corresponding to the $r$-th ray {of cluster $c$}, and is drawn from a zero-mean-unit-variance circular symmetric Gaussian distribution, {$\mathbf{a}_{\mathrm{Tx}}$} denotes the response of the overall TX array, while {${a}_{\mathrm{Rx}}$ accounts for the phase between the clusters and the users}. {In the special case of uniform linear array (ULA) of size $LM$, the TX array vector reads $\mathbf{a}_{\mathrm{Tx}}(\zeta)= [1,  e^{j{2\pi}\Delta_\mathrm{ant} \mathrm{sin}(\zeta)}, \dots, e^{j(LM-1){2\pi}\Delta_\mathrm{ant} \mathrm{sin}(\zeta)}]^T$} where $\Delta_\mathrm{ant}$ is the antenna spacing relative to the wavelength.

The corresponding delay-$d$ multiuser MIMO channel matrix reads then $\mathbf{H}[d] = (\mathbf{h}_1[d], \mathbf{h}_2[d], \dots, \mathbf{h}_U[d])^{T} \in \mathbb{C}^{U\times {LM}}$. 
Finally, the corresponding multiuser frequency-domain response at subcarrier $k$, denoted by {$\boldsymbol{\Gamma}[k] =  (\boldsymbol{\gamma}_1[k], \boldsymbol{\gamma}_2[k], \dots, \boldsymbol{\gamma}_U[k])^{T}\in \mathbb{C}^{U\times LM}$}, is given by
\begin{equation}
    {\boldsymbol{\Gamma}[k]} = \sum_{\substack{d=0}}^{D-1}\mathbf{H}[d]e^{-j\frac{2\pi kd}{K_{\mathrm{FFT}}}}
    \label{Chan Freq}
\end{equation}
where $D$ denotes the assumed channel impulse response length in samples. A LOS component can also be added, on top of the channel model in (\ref{channel_model}), in order to account for {Rician} fading with any given {Rician} K-factor defined as the power ratio between the  received LOS and NLOS components \cite{propagation_book}. {Finally, it is noted that a similar channel modeling approach is also utilized in 3GPP mobile radio standardization \cite{3GPPTR38901}}.

\subsection{Design of Digital and Analog Precoders}\label{Design_Precoders}
The design and optimization of the digital and analog precoders in hybrid MIMO transmitters is generally a challenging problem. The analog and digital precoders constitute a cascaded system, therefore, both blocks are coupled making the resulting optimization problem non-convex  \cite{HybridPrecoding1,HybridPrecoding2,Intro_3,Precoding1}. {Furthermore, since the analog precoders are typically implemented as a network of phase shifters, this imposes additional constraints, such as having a limited set of available phase rotations.} One common approach is thus to decouple the design of the baseband and analog precoders \cite{HybridPrecoding2}. The analog precoder can be first selected based on beamsteering the signals towards the dominant directions of the channel, while the BB precoding, that acts over the equivalent beamformed channel (analog precoder and actual multi-user MIMO channel), is responsible for reducing the multi-user interference and compensating for the frequency-selectivity of the channel. {It is noted that the optimization of the precoders is a large topic of its own, while the fairly straight-forward decoupling based approach is considered in this article.}
Provided that the analog precoder $\mathbf{W}$ is known or fixed, the BB precoding matrix at the $k$-th subcarrier can be obtained in a straight-forward manner, by utilizing the equivalent channel matrix { $\boldsymbol{\Gamma}_{\mathrm{eq}}[k] = \boldsymbol{\Gamma}[k]\mathbf{W}$.} For example, the zero-forcing (ZF) and regularized ZF (RZF)  precoders essentially read \cite{Traditional_MIMO2,RZF}
\begin{align}
    \mathbf{F}_{\mathrm{ZF}}[k] &= \boldsymbol{\Gamma}^H_{\mathrm{eq}}[k](\boldsymbol{\Gamma}_{\mathrm{eq}}[k]\boldsymbol{\Gamma}_{\mathrm{eq}}^H[k])^{-1}\\
    \mathbf{F}_{\mathrm{RZF}}[k] &= \boldsymbol{\Gamma}^H_{\mathrm{eq}}[k](\boldsymbol{\Gamma}_{\mathrm{eq}}[k]\boldsymbol{\Gamma}_{\mathrm{eq}}^H[k] + \delta I)^{-1}.
\end{align}
For transmit power normalization, additional scaling factors can be introduced, building on, e.g., a sum-power constraint \cite{AE1,HybridPrecoding1,Reduced_HybridMIMO}. {In practice, the equivalent channels are known only with finite accuracy, an aspect that we address through our numerical experiments in Section VII.}

For the reduced-complexity architecture, the composite analog precoder matrix is in general of the form 
\begin{equation}
\mathbf{W} = 
\mathbf{I}_M \hspace{0.1mm} \odot \hspace{0.1mm} (\mathbf{w}_1, \hdots, \mathbf{w}_L) 
\end{equation}
where $\mathbf{I}_M$ is an $M \times M$ identify matrix, the operator  
$\odot$ refers to Khatri-Rao product {\cite{khatri_rao}}, $\mathbf{w}_l = (w_{l,1},w_{l,2}, \dots, w_{l,M})^T \in \mathbb{C}^{M \times 1}$ with $|w_{l,m}| = 1 $ $\forall l,m$ is the beamforming vector of the $l$-th subarray. Interestingly, the weights $w_{l,m}$ can be optimized in multiple ways, while in this work we conceptually differentiate between the following two main alternatives:
\subsubsection{Single-beam analog beamformer} A subarray generates a single beam towards the main channel tap of a particular user. An individual user is then being primarily served by a single subarray. It is, however, important to note that the actual received signal of every user is still contributed by the transmitted signals of all the subarrays since practical beampatterns provide only limited spatial isolation. 
\subsubsection{Multi-beam analog beamformer} Each subarray generates multiple beams, one per user, simultaneously. All the users are then more evenly served by all the subarrays, and thus the received signals are not dominated by the transmissions from a single subarray. 
In order to generate multiple simultaneous beams through phase-only precoding, one can refer, e.g., to \cite{AnalogBeamforming} and \cite{Heath_multibeam}. 

It is noted that in general, the multi-beam analog beamforming approach can, e.g., facilitate additional macro-diversity, if the different sub-arrays are located over larger distances, while also offering more degrees of freedom for frequency multiplexing. Additionally, as noted in \cite{Heath_multibeam}, multiple simultaneous beams can also facilitate joint communications and sensing/radar. In this article, we consider both alternative approaches in a comparative manner, while do not explicitly claim one better than the other. 

\section{Modeling and Analysis of PA Induced Nonlinear Distortion}\label{Nonlinear_model}
To build the basis for linearization, the modeling of the PA-induced nonlinear distortion is next pursued. {To account for wideband memory effects in practical mmWave PAs, we consider memory polynomial based PA models in the analysis}. Additionally, different PA units are assumed mutually different, no DPD processing is yet considered in this section, {i.e., $\tilde{x}(n) = x(n)$ in the context of Fig. 1}, and all modeling is carried out in discrete-time baseband equivalent domain. The extension to the challenging crosstalk case is provided separately, for presentation convenience, in Section \ref{sec:crosstalk}. 

Consider the $m$th antenna branch in the $l$th subarray, and let $v_{l,m}(n) = w_{l,m}x_l(n)$ denote the PA input signal, where $w_{l,m}$ refers to the analog beamformer weight while $x_l(n)$ denotes the digitally precoded sample sequence of the $l$th TX. The corresponding PA output signal can then be expressed as
\begin{equation}
\begin{split}
y_{l,m}(n) &= \sum_{\substack{p=1 \\ p, \text{odd}}}^{P} \alpha_{l,m,p}(n)\star v_{l,m}(n)|v_{l,m}(n)|^{p-1}\\
 &= w_{l,m}\sum_{\substack{p=1 \\ p, \text{odd}}}^{P} \alpha_{l,m,p}(n) \star x_l(n)|w_{l,m}x_l(n)|^{p-1}, \label{eq:PA_out_m}
\end{split}
\end{equation}
where $\alpha_{l,m,p}(n)$ stands for the $p$th order PA impulse response at the $m$th antenna branch of the subarray $l$, while $P$ is the corresponding polynomial order and $\star$ is the discrete-time convolution operator. Since $|w_{l,m}| = 1$, the PA output signal in (\ref{eq:PA_out_m}) can be re-written as
\begin{align}
    y_{l,m}(n) &= w_{l,m}\sum_{\substack{p=1 \\ p, \text{odd}}}^{P} \alpha_{l,m,p}(n)\star\psi_{l,p}(n), \label{eq:PA_out_m2}
\end{align}
where $\psi_{l,p}(n) = x_l(n)|x_l(n)|^{p-1}$ denotes the so-called static nonlinear (SNL) basis function of order $p$. 

Let us next consider the observable combined signal at user $u$, being contributed by all antenna elements of all subarrays. Denoting the impulse response between the $m$th antenna element of the $l$th subarray and the $u$th user by $h_{l,m,u}(n)$ -- {which corresponds to the sequence of complex numbers at the row of the matrix $(\mathbf{h}_u[0], \hdots, \mathbf{h}_u[D-1])$ defined by the considered antenna index 
 --} the received signal excluding additive thermal noise for notational simplicity reads 
\begin{equation}
    z_u(n) = \sum_{\substack{l=1}}^{L}\sum_{\substack{m=1}}^{M} w_{l,m}h_{l,m,u}(n)\star\sum_{\substack{p=1 \\ p, \text{odd}}}^{P} \alpha_{l,m,p}(n)\star\psi_{l,p}(n). \label{RX_signal1}
\end{equation}
  
 Assuming next that the individual channels within a single subarray are clearly correlated, {a common assumption at mmWaves}, one can argue that $h_{l,m,u}(n) \approx h_{l,u}(n)e^{j\beta^l_{m,u}}$, and thus rewrite (\ref{RX_signal1}) as 
 \begin{equation}
    z_u(n) \approx \sum_{\substack{l=1}}^{L}h_{l,u}(n)\star\sum_{\substack{m=1}}^{M}w_{l,m}e^{j\beta^l_{m,u}}\sum_{\substack{p=1 \\ p, \text{odd}}}^{P} \alpha_{l,m,p}(n) \star \psi_{l,p}(n) \label{RX_signal_corr}
\end{equation}
\normalsize
where $e^{j\beta^l_{m,u}}$ stems from the phase differences between the signals due to the array geometry as well as exact propagation conditions. 
{Such an approximation is well-argued at mmWaves, where there is typically a dominating LOS path and only few scatterers \cite{Precoding1,Reduced_HybridMIMO}. The assumption naturally holds also under pure LOS scenario, as well as under geometric channel models with small antenna spacing such that the spatial correlation is high \cite{ChannelCorr}}. It is also noted that for notational convenience, the phase of the dominant channel tap of  $h_{l,u}(n)$ is assumed to be embedded in $e^{j\beta^l_{m,u}}$. {Finally, it is noted that no assumptions regarding the correlation of channels between subarrays are made, and that in numerical evaluations in Section~\ref{Results}, array propagation and received signals are modeled and calculated without any approximations.}

In order to have a better insight into the structure of the observable nonlinear distortion, we focus next on the received signals of two users, say $u$ and $u'$, and specifically investigate the contribution of the $l$th TX chain only, expressed as 
\begin{align}
    z^l_u(n) &= h_{l,u}(n)\star\sum_{\substack{m=1}}^{M}\sum_{\substack{p=1 \\ p, \text{odd}}}^{P} e^{j\beta^l_{m,u}}w_{l,m}\alpha_{l,m,p}(n)\star \psi_{l,p}(n) \label{subarray1}  \\
    z^l_{u'}(n) &= h_{l,u'}(n)\star\sum_{\substack{m=1}}^{M}\sum_{\substack{p=1 \\ p, \text{odd}}}^{P} e^{j\beta^l_{m,u'}}w_{l,m}\alpha_{l,m,p}(n)\star \psi_{l,p}(n) \label{subarray2} 
\end{align}
It can be seen from (\ref{subarray1}) and (\ref{subarray2}) that the received signals at different receivers, stemming from a given subarray, have a very similar structure. The nonlinear terms are shaped by the same analog precoder coefficients and the same PA responses, while only the channel impulse responses and the element-wise phase differences differ. Then, by considering the multi-beam analog beamformer discussed in Section \ref{Design_Precoders}, for generality purposes and to harness true multi-user hybrid MIMO, coherent combining towards both users is achieved, {implying that $\sum_{\substack{m=1}}^{M}e^{j\beta^l_{m,u}}w_{l,m}\alpha_{l,m,p}(n)$ and $\sum_{\substack{m=1}}^{M}e^{j\beta^l_{m,u'}}w_{l,m}\alpha_{l,m,p}(n)$ are mutually very similar, for given $l$ and $p$, up to a possible constant phase difference that is independent of $p$. More specifically, it can be shown  that $\sum_{\substack{m=1}}^{M}e^{j\beta^l_{m,u}}w_{l,m}\alpha_{l,m,p}(n) \approx e^{j\xi^l_{u,u'}}(\sum_{\substack{m=1}}^{M}e^{j\beta^l_{m,u'}}w_{l,m}\alpha_{l,m,p}(n))$, where the phase $\xi^l_{u,u'}$ does not depend on $p$.} {This is because with practical PA implementation technologies and manufacturing processes, it is primarily the implementation inaccuracies that constitute to the mutual differences of the PA samples.} Hence, (\ref{subarray1}) and (\ref{subarray2}) can be essentially re-written as
 \begin{align}
    z^l_u(n) &= {h^{\text{eff}}_{l,u}(n)}\star\sum_{\substack{p=1 \\ p, \text{odd}}}^{P} \alpha^{\mathrm{tot}}_{l,p}(n)\star \psi_{l,p}(n) \label{coherent_combination1}\\
    z^l_{u'}(n) &= {h^{\text{eff}}_{l,u'}(n)}\star\sum_{\substack{p=1 \\ p, \text{odd}}}^{P} \alpha^{\mathrm{tot}}_{l,p}(n)\star \psi_{l,p}(n),\label{coherent_combination2}
\end{align}
where $\alpha^{\mathrm{tot}}_{l,p}(n) = \sum_{\substack{m=1}}^{M}\alpha_{l,m,p}(n)$ stands for the equivalent $p$-th order PA {impulse response of the whole subarray, while the possible additional user-specific constant scalar that is common for all values of $p$ is lumped to the effective channel impulse response {$h^{\text{eff}}_{l,u}(n)$}}. {The above approximation holds very accurately with all the measured PA models shared along the article.}

As acknowledged already in \cite{OOB_Mollen,OOB_Emissions_Ours}, the linear and nonlinear signal terms get beamformed towards the same directions. This is clearly visible already in (\ref{subarray1}) and (\ref{subarray2}), since the nonlinear basis functions are subject to similar effective beamforming gains of the form $\sum_{\substack{m=1}}^{M}e^{j\beta^l_{m,u}}w_{l,m}\alpha_{l,m,p}(n)$. Therefore, when multi-beam analog beamformers are adopted in different subarrays, there are as many harmful directions for the distortion, per subarray, as there are intended users. However, very importantly, it can also be observed that apart from the linear filtering effect, the signals in (\ref{coherent_combination1}) and (\ref{coherent_combination2}) are both basically identical memory polynomials of the original digital signal samples $x_l(n)$, expressed through the SNL basis functions $\psi_{l,p}(n)$ and the effective or equivalent PA filter coefficients of the whole subarray. Thus, the observable nonlinear distortion at the two considered receivers, contributed by one subarray, is essentially the same, except for the channel filtering, and can be thus modeled with the same {memory} polynomial. {This is one of the key results, as it implies that a single DPD per subarray can simultaneously provide linearization towards all intended receivers -- something that is essential, since the nonlinear distortion from an individual subarray is strongest towards these directions due to multi-beam beamforming. This modeling and analysis thus form the technical basis for the adoption of the single-input DPD system and parameter learning principles described in the next section.} 

\section{Single-Input DPD System and Parameter Learning Solution}\label{DPD structure}
{Based on the above nonlinear distortion analysis, we now proceed by adopting single-input DPD processing, structurally similar to  that considered in \cite{DPD_MM_5} in single phased-array context, in the different parallel transmit paths, and analytically show that the observable distortion can be suppressed.}

\subsection{DPD Processing and Observable Distortion Suppression}

Motivated by (\ref{coherent_combination1}) and (\ref{coherent_combination2}), and their generalization to $U$ users, we argue that a single {memory}  polynomial DPD can model and suppress the nonlinear distortion stemming from the corresponding subarray towards all intended receivers.
Thus, the core DPD engine {\cite{DPD_MM_5}} in the $l$th TX path is expressed as 
\begin{align}
\tilde{x}_{l}(n) =  x_{l}(n) + \sum_{\substack{q=3\\ q, \text{odd}}}^{Q} \lambda^*_{l,q}(n)\star\psi_{l,q}(n).
\label{eq:PA_In_with_DPD}
\end{align}
where $\psi_{l,q}(n)$, $q = 3,5,\dots Q$, denote the DPD basis functions up to order $Q$, while $\lambda_{l,q}(n)$, $q = 3,5,\dots Q$  denote the corresponding DPD filter {impulse responses}. We have deliberately excluded processing the amplitude and phase of the linear term in (\ref{eq:PA_In_with_DPD}), as our main purpose is to suppress the nonlinear distortion while linear response equalization is anyway pursued separately in the RX side. 

Assuming that the above type of DPD processing is executed in every TX path, we will next explicitly show that the total observable nonlinear distortion can be efficiently suppressed as long as the DPD filters are properly optimized. To this end, we substitute the DPD output signals in 
(\ref{eq:PA_In_with_DPD}), for $l=1,2, \dots, L$, as the PA input signals in the {basis functions in (\ref{coherent_combination1})}, which together with summing over $L$ yields  
\begin{equation}
\begin{split}
    \tilde{z}_u(n) &=\sum_{\substack{l=1}}^{L}{h^{\text{eff}}_{l,u}}(n)\star \alpha^{\text{tot}}_{l,1}(n)\star  \psi_{l,1}(n) \\
    &+ \sum_{\substack{l=1}}^{L}{h^{\text{eff}}_{l,u}(n)}\star \sum_{\substack{q=3\\ q, \text{odd}}}^{Q}  \lambda^*_{l,q}(n)\star\alpha^{\text{tot}}_{l,1}(n)\star \psi_{l,q}(n) \\
    &+ \sum_{\substack{l=1}}^{L}{h^{\text{eff}}_{l,u}(n)}\star\sum_{\substack{p=3 \\ p, \text{odd}}}^{P} \alpha^{\text{tot}}_{l,p}(n)\star  \psi_{l,p}(n), \label{lin_nonlin_expanded}
\end{split}
\end{equation}
In above, the first line corresponds to the linear signal while the rest are nonlinear terms. In reaching the above expression it was further assumed that the nonlinear terms introduced by the DPD in (\ref{eq:PA_In_with_DPD})  are clearly weaker than the linear signal -- an assumption that essentially holds in practice -- and hence themselves only excite the linear responses of the PAs. 

For notational simplicity, we next further assume that the DPD nonlinearity order $Q$ is equal to the PA nonlinearity order $P$, which allows us to rewrite (\ref{lin_nonlin_expanded}) as
\begin{equation}
\begin{split}
    \tilde{z}_u(n) &= \sum_{\substack{l=1}}^{L}{h^{\text{eff}}_{l,u}(n)}\star \alpha^{\text{tot}}_{l,1}(n)\star \psi_{l,1}(n) + \sum_{\substack{l=1}}^{L}{h^{\text{eff}}_{l,u}(n)}\\
    &  \star\sum_{\substack{p=3\\ p, \text{odd}}}^{P} \big(\lambda^*_{l,p}(n)\star\alpha^{\text{tot}}_{l,1}(n) + \alpha^{\text{tot}}_{l,p}(n)\big)\star \psi_{l,p}(n). \label{RX_DPD2}
\end{split}
\end{equation}
Based on (\ref{RX_DPD2}), one can explicitly see that the DPD filters $\lambda_{l,p}(n)$ can be chosen such that the nonlinear distortion at the receiver end is suppressed, i.e., $\lambda^*_{l,p}(n)\star\alpha^{\mathrm{tot}}_{l,1}(n) + \alpha^{\mathrm{tot}}_{l,p}(n)=0$ {for the considered range of lags}. This thus more formally shows that $L$ {memory} polynomial DPDs, one per subarray, can effectively linearize $L \times M$ different PAs, particularly when considering the observable linear distortion at RX side, despite all the PA units being generally mutually different. Importantly, the expression in (\ref{RX_DPD2}) also indicates that DPD filters that yield good nonlinear distortion suppression are independent of the actual channel realization. Thus, while the beamforming coefficients should obviously follow the changes in the channel characteristics, the DPD system needs to track changes only in the PAs, {specifically when there is no substantial crosstalk like assumed in this section}. This will be also verified and demonstrated through the numerical experiments, {while the challenging crosstalk scenario and its implications are addressed in Section~\ref{sec:crosstalk}}.

{Importantly, the expression in (\ref{RX_DPD2}) also reveals that the multipath channel filtering is essentially common to all nonlinear distortion terms, stemming from an individual subarray, while the term on the second line is identical to coherent sum of the PA output signals of the corresponding subarray. Thus, instead of seeking to use true OTA signals for DPD learning -- an option discussed, e.g., in \cite{Swedes_review} and the references therein -- hardware based combining illustrated in Fig. \ref{fig:WholeSystem} can reproduce coherent summation of the PA output signals, and thus facilitate efficient DPD learning. 
}

\subsection{Combined Feedback based DPD Learning}
In practice, the nonlinear responses of the individual PA units are unknown and can also change over time. Thus, proper parameter learning is needed. Building on the above modeling {and adopting a similar approach as done in \cite{DPD_MM_5} in the single phased-array context}, the considered DPD parameter learning builds on the coherently combined observations of the subarray signals as shown already in Fig. \ref{fig:WholeSystem}. To this end, and considering the PA output signals in (\ref{eq:PA_out_m2}), the baseband combined feedback signal in the $l$-th TX or subarray reads
\begin{align}
    z^l_{\mathrm{fb}}(n) &= \sum_{\substack{m=1}}^{M}w_{l,m}^*y_{l,m}(n) \\
    & =\sum_{\substack{m=1}}^{M} |w_{l,m}|^2\sum_{\substack{p=1 \\ p, \text{odd}}}^{P} \alpha_{l,m,p}(n)\star \psi_{l,p}(n) \\
    & = \sum_{\substack{p=1 \\ p, \text{odd}}}^{P} \alpha^{\mathrm{tot}}_{l,p}(n)\star \psi_{l,p}(n).\label{learning_feedback_signal}
\end{align}
As can be observed, the combined feedback signal is structurally identical to the actual received signal model in (\ref{coherent_combination1}), except for the channel filtering effect, forming thus good basis for DPD filter optimization.  

{Regarding the feedback implementation, we note that in time division duplex systems, the receiver-side beamformer of the overall transceiver (e.g. the 5G NR base-station) could potentially be re-purposed for the feedback beamformer task, to alleviate the hardware complexity.} {We also note that the true OTA combined received signals can, potentially, be used for the parameter learning, as discussed at general level in \cite{Swedes_review} and references therein. Such an approach would simplify the transmitter system, in terms of the feedback couplers, but would also call for specific OTA measurement receivers and some mechanism to provide the measured I/Q signal samples to the transmitter system. However, the potential load-modulation dependency on the steering angle \cite{Swedes_review} in practical transmitters would make this method inaccurate or difficult to implement. This is further discussed in the following sections. Additionally, the frequency-selectivity of the broadband channels could potentially hinder the DPD parameter learning, in the sense that the DPD system would interpret the channel memory as part of the PAs' memory. Hence, in this work, we assume the hardware combiner based feedback approach.}

Generally-speaking the feedback model in (\ref{learning_feedback_signal}) allows for multiple alternative approaches for DPD learning. One option is to do direct least-squares (LS) estimation of the effective impulse responses $\alpha^{\mathrm{tot}}_{l,p}(n)$, and then use these estimates together with (\ref{RX_DPD2}) to solve for the DPD filters $\lambda_{l,p}(n)$ through $\lambda^*_{l,p}(n) \star \alpha^{\mathrm{tot}}_{l,1}(n) + \alpha^{\mathrm{tot}}_{l,p}(n)=0$. Another alternative would be to deploy indirect learning architecture (ILA) \cite{Trad_DPD} where the combined feedback signal in (\ref{learning_feedback_signal}) is fed into a memory polynomial post-distorter whose coefficients are estimated through, e.g., LS, and then substituted as an actual predistorter.  

{In this article, inspired by our earlier work in \cite{DPD_MM_5} in the context of single-user MIMO, we pursue CL adaptive learning solutions through iterative gradient methods. Such CL approach allows for implementation-feasible parameter tracking, and is also shown in the next section to be robust against the potential crosstalk.} Specifically, the learning system seeks to minimize the observable nonlinear distortion by minimizing the correlation between the  nonlinear distortion in the combined feedback signal and the DPD SNL basis functions $\psi_{l,q}(n)$, $q = 3,5,\dots Q$, {over the considered lags in the DPD filters}. 
Such learning procedure is carried out in parallel in all $L$ transmitters. To extract the effective nonlinear distortion in the combined feedback signal $z^l_{\mathrm{fb}}(n)$, we assume that an estimate of the effective linear impulse response {$g_{l}(n)={\alpha}^{\text{tot}}_{l,1}(n)$}, denoted by $\hat{g_l}(n)$, is available. Based on this, the effective nonlinear distortion can be extracted as
\begin{equation}
    e_l(n) = z^l_{\mathrm{fb}}(n) - \hat{g_l}(n) \star x_l(n).\label{error_signal}
\end{equation}
In practice, $\hat{g_l}(n)$ can be obtained, e.g., by means of block LS {or any other linear estimator, by using the signal of the $l$th observation receiver and the corresponding known transmit data $x_l(n)$. Assuming that the feedback combiner weights are well-matched to the corresponding main path RF beamformer, the effective linear loop response ${g_l}(n)$ is not beam-dependent, and thus can be estimated infrequently.} 

The exact computing algorithm, seeking to tune the DPD coefficients to decorrelate the feedback nonlinear distortion or error signal $e_l(n)$ and the SNL basis functions can build on, e.g., well-known block-LMS \cite{Haykin_AdaptiveFilters}, and reads 
{\begin{equation}
    \boldsymbol{\lambda}_l(i+1) = \boldsymbol{\lambda}_l(i) - \mu_l\mathbf{R}^{-1}\boldsymbol{\Psi}_l^T(i)\mathbf{e}_l^*(i),\label{eq:learning_rule}
\end{equation}
where $i$ denotes the learning block index, $\boldsymbol{\lambda}_l(i)$ is the vector containing the $l$th DPD filter coefficients, $\mu_l$ denotes the learning rate, while $\mathbf{R} = \mathbb{E}\{\boldsymbol{\psi}(n)\boldsymbol{\psi}(n)^H\}$ is the covariance matrix of the DPD input vector, which can be precomputed. Furthermore,   $\boldsymbol{\Psi}_l(i)$ contains the considered basis function samples properly stacked into a matrix over the learning block size, while $\mathbf{e}_l(i)$ is the corresponding error signal vector.
As the basis functions are in practice mutually correlated, the self-orthogonalized version of the classical block-LMS, as shown in (\ref{eq:learning_rule}), is adopted in this work to ensure smooth and fast convergence. Alternatively, one could consider to orthogonalize the basis functions matrix, e.g., through Cholesky decomposition. However, the self-orthogonalizing rule is preferred, complexity-wise, since in such approach the orthogonalization transformation is only executed in the parameter learning loop, while the DPD main path processing can utilize the original basis function samples.
}

{\section{Extension to Transmitter Systems with Crosstalk}\label{sec:crosstalk}
Due to the closely spaced antenna units and high level of integration in practical active antenna array transmitters, crosstalk is inevitable \cite{Swedes_review,28GHz_crosstalk}. In this section, we thus provide an important extension of the previous developments to hybrid MIMO transmitter systems affected by such crosstalk.
{Since the PA units of different subarrays or TX chains are commonly implemented on separate integrated circuits, crosstalk prior to the PAs is assumed to take place only within the individual subarrays, as separate ICs imply higher physical isolation. On the other hand, antenna crosstalk can basically occur between any two antennas. However, as the crosstalk level decays as a function of the physical distance \cite{Prediction_distortion,28GHz_crosstalk}, antenna crosstalk from different subarrays is in general weaker, and is modelled linearly in this work, similar to \cite{MIMO_DPD_2,MIMO_DPD_3}. It is also noted that while the so-called dual-input model in \cite{Prediction_distortion,Crosstalk_Chalmers} is a more physically-inspired model of the load modulation phenomenon, the PA input/output crosstalk modeling approach adopted in this work provides, under the above assumptions, the same nonlinear distortion products while is also reflecting the nonlinear distortion beam-dependency}.

\subsection{Observable Nonlinear Distortion Under Crosstalk}
First, to account for PA input crosstalk \cite{MIMO_DPD_2,MIMO_DPD_3}, an individual PA input signal is re-defined as a linear combination of all the phase-shifted signals of the corresponding sub-array, and thus reads
\begin{align}
    \bar{v}_{l,m}(n) &= \Big(w_{l,m} + \sum_{\substack{i=1\\i \neq m}}^{M}b_{l,i,m}w_{l,i}\Big)x_l(n) \\
    &= \bar{w}_{l,m}x_l(n),
\end{align}
where $b_{l,i,m}$ is the complex crosstalk or coupling factor between the $i$th and the $m$th PA inputs in subarray $l$, and $\bar{w}_{l,m} = w_{l,m} + \sum_{\substack{i=1\\i \neq m}}^{M}b_{l,i,m}w_{l,i}$ is the corresponding effective analog beamforming coefficient. In general, it can be noted that $|\bar{w}_{l,m}|\neq 1$, which imposes in general a dependency of the PA nonlinear behaviour on the steering angle, similar to load modulation \cite{Swedes_review}. Such beam-dependent nonlinear behaviour can potentially call for continuous DPD learning, in real time, in order to adapt to and track the changing nonlinear distortion behavior. This is demonstrated and investigated further in Section \ref{Results} through numerical examples.

Then, in the presence of antenna coupling, the effective PA output or antenna signal is contributed by the signals leaking from all the other antennas, consequently, (\ref{eq:PA_out_m2}) is re-written as 
\begin{equation}
\begin{split}
   \bar{y}_{l,m}(n) &= \bar{w}_{l,m}\sum_{\substack{p=1 \\ p, \text{odd}}}^{P}\bar{\alpha}_{l,m,p}(n)\star\psi_{l,p}(n)\\
   &+ \sum_{\substack{k=1}}^{L}\sum_{\substack{i=1\\i \neq m}}^{M}c_{k,l,m,i}\bar{w}_{k,i}\sum_{\substack{p=1 \\ p, \text{odd}}}^{P}\bar{\alpha}_{k,i,p}(n)\star\psi_{k,p}(n)\label{eq:PA_output_crosstalk},
\end{split}
\end{equation}
where $\bar{\alpha}_{l,m,p}(n)= |\bar{w}_{l,m}|^{p-1}\alpha_{l,m,p}(n)$ and $c_{k,l,i,m}$ is the antenna complex coupling factor from the $i$th to the $m$th antenna between the subarrays $k$ and $l$. 

\begin{table*}[]
\centering
\caption{\textsc{{DPD main path processing complexity per sample and total DPD learning complexity of the hybrid MIMO transmitter. $T=1$ if there is no crosstalk, and $T=L$ if crosstalk is considered.}}}
\label{tab:complexity}
\begin{tabular}{cccc}\hline & & & \vspace{-2.5mm} \\ 
                  &  & \textbf{CL DPD} & \textbf{ILA DPD} \\ \hline  & & & \vspace{-1mm}\\
\multirow{2}{*}{\textbf{DPD}} & BF gen.& ${L(2N_{\text{IBF}}-1)}$  & ${L(2N_{\text{IBF}}-1)}$ \vspace{1mm} \\ 
                  & Filtering &  ${L(8N_{\text{BF}}-8)}$ & ${L(8N_{\text{BF}}-2)}$  \\ & & & \vspace{-.0mm}\\
\multirow{2}{*}{\vspace{-2mm}\textbf{Learning}} & BF gen. &  $-$ & ${LI_{\text{ILA}}(2N_{\text{IBF}}-1)N_{\text{ILA}}}$  \\
                  & DPD est. &  \makecell{ ${LI_{\text{CL}}(8N_{\text{BF}}N_{\text{CL}}+8N_{\text{BF}}^2)} + \ceil*{4LT(N_{\text{CL}} + \frac{D_{\text{lin}}}{3})(D_{\text{lin}})^2}$\\ $+LT(8D_{\text{lin}}-2)N_{\text{CL}}I_{\text{CL}}+2LN_{\text{CL}}I_{\text{CL}}$}  & $\ceil*{4L(I_\text{ILA}(N_{\text{ILA}}+\frac{N_{\text{BF}}}{3})(N_{\text{BF}})^2}$  \\ \cline{1-4} 
\end{tabular}
\end{table*}

The corresponding received or observable signal by user $u$ can now be shown to read
 \begin{equation}
 \begin{split}
     \bar{z}_u(n)&=\sum_{\substack{l=1}}^{L} {h^{\text{eff}}_{l,u}(n)}\star\\
     &\Big(\sum_{\substack{p=1 \\ p, \text{odd}}}^{P}\bar{\alpha}^{\text{tot}}_{l,p}(n)\star\psi_{l,p}(n) +   \sum_{\substack{k=1 \\ k\neq l }}^{L}\sum_{\substack{p=1 \\ p, \text{odd}}}^{P}{f}^{\text{tot}}_{l,k,p}(n)\star\psi_{k,p}(n)\Big)\label{eq:rx_crosstalk}
 \end{split}{}
 \end{equation}
 where $\bar{\alpha}^{\text{tot}}_{l,p}(n) = \sum_{\substack{m=1}}^{M}\Big(e^{j\Delta\beta^{l,l}_{m,m}} + \sum_{\substack{i=1 \\ i \neq m}}^{M} e^{j\Delta\beta_{i,m}}b_{i,m}+\sum_{\substack{i=1\\ i\neq m }}^{M}c_{l,l,i,m} e^{j\Delta{\beta}^{k,l}_{i,m}}\Big) 
\bar{\alpha}_{l,m,p}(n)$, $\Delta\beta^{k,l}_{i,m} = \beta^k_{i}+\bar{\beta}^{l}_{m}$, where $\bar{\beta}^{l}_{m}$ is the phase of the corresponding analog beamformer coefficient, and ${f}^{\text{tot}}_{l,k,p}(n) = \sum_{\substack{m=1}}^{M}\sum_{\substack{i=1\\ i\neq m }}^{M}c_{k,l,i,m} e^{j\Delta{\beta}^{k,l}_{i,m}}\bar{\alpha}_{k,i,p}(n)$. The detailed analysis steps and derivation of (\ref{eq:rx_crosstalk}), starting from (\ref{eq:PA_output_crosstalk}), are provided in the  Appendix \ref{sec:Appendix}.

Similarly, the output of the $l$th observation receiver, given originally in (\ref{learning_feedback_signal}), can now be re-expressed under crosstalk as 
\begin{align}
    \bar{z}^l_{\mathrm{fb}}(n) 
        &= \sum_{\substack{p=1 \\ p, \text{odd}}}^{P}\bar{\alpha}^{\text{tot}}_{l,p}(n)\star\psi_{l,p}(n) +   \sum_{\substack{k=1 \\ k\neq l }}^{L}\sum_{\substack{p=1 \\ p, \text{odd}}}^{P}{f}^{\text{tot}}_{l,k,p}(n)\star\psi_{k,p}(n)
     \label{eq:z_pre}
\end{align}
which has the same structure as the true received signal except for the actual channel filtering. Additionally, the same derivation steps described in Appendix \ref{sec:Appendix} for (\ref{eq:rx_crosstalk}) also apply here, to reach the final form in (\ref{eq:z_pre}).

\subsection{Single-Input Closed-Loop DPD Under Crosstalk}
Essentially, the models in (\ref{eq:rx_crosstalk}) and (\ref{eq:z_pre}) contain $L$ memory polynomials (MPs). Thus, one approach to linearize is to adopt $L$ memory polynomials, in parallel, in each TX branch, which means essentially a multi-input DPD unit.
However, as the expression for $f^{\text{tot}}_{k,p}(n), k\neq l$ below (\ref{eq:rx_crosstalk}) shows, the $L-1$ MP terms do not generally combine coherently towards the receivers. Hence, to emphasize implementation-feasible processing complexity, we still pursue linearizing the transmitter system with the individual single-input MP based DPD units, one per TX branch, as described in (14).

Next, by denoting the linear responses ($p=1$) as $\bar{\alpha}^{\text{tot}}_{l,1}(n)=g_{l,l}(n)$ and $f^{\text{tot}}_{l,k,1}(n)=g_{k,l}(n),k\neq l$, the observable feedback signal in (\ref{eq:z_pre}) can be re-written as
\begin{equation}
    \bar{z}^l_{\text{fb}}(n)= \sum_{\substack{k=1}}^{L}g_{k,l}(n)\star x_k(n) + d_l(n),
    \label{eq:z_fb_crosstalk}
\end{equation}
where $g_{k,l}(n)$ is the effective linear impulse response from the input of the $k$th TX chain to the output of the $l$th observation receiver, while the total effective nonlinear distortion is lumped in $d_l(n)$.
The error signal for the CL parameter learning system is then given by
\begin{equation}
\bar{e}_l(n) = \bar{z}^l_{\mathrm{fb}}(n) - \sum_{\substack{k=1}}^{L}\hat{g}_{k,l}(n) \star x_k(n).
\end{equation}
where the second-term seeks to suppress the linear contribution of each transmitter from a given observation receiver such that an estimate of the prevailing nonlinear distortion is obtained. The actual learning rule follows then (21). Again, similar to the discussion along (20), the estimates of the linear impulse responses can be obtained through, e.g., linear least-squares {or any other linear estimator, by using the known transmit signals and the signals of the observation receivers. It is also noted that the coupling responses $g_{k,l}(n)$, $k \neq l$, are, strictly-speaking, beam-dependent since they reflect OTA coupling from $k$th TX to $l$th observation RX, and thus stem from non-matched main-path beamformer and feedback anti-beamformer. However, it is straight-forward to show that for given $l$ and $k$, such beam-dependence corresponds to a complex scalar multiplier that can be calculated based on the corresponding beamformer and anti-beamformer weights. Hence, the actual estimation procedures can be executed based on only the potential time-variance of the true physical coupling features.} 

Compared to widely-applied indirect learning, which is known to largely suffer from the crosstalk \cite{MIMO_DPD_2,MIMO_DPD_3,MIMO_DPD_4} in single-input DPD context, the CL system seeking to decorrelate the prevailing error signal and the single-input DPD basis functions provides inherent robustness against the crosstalk. This is because in the CL system, the basis functions utilized in the parameter learning are calculated from the clean digital signal samples (i.e., $x_l(n)$) while in ILA-based learning the post-distorter basis functions are calculated using the feedback samples (i.e., $\bar{z}^l_\text{fb}(n)$) which suffer from crosstalk. These are demonstrated through simulations in Section VII-D. }

\section{Computational Complexity Analysis}\label{sec:complexity}
{In this section, a complexity analysis of the CL DPD solution with self-orthogonalized learning and the widely adopted ILA-based DPD with LS fitting for parameter learning is conducted in the presence and absence of crosstalk. We differentiate between the complexity of the DPD learning, and the complexity of the actual main path linearization. The complexity of the main path is usually more critical as it is to be executed in real time along with the data transmission, while the parameter learning is executed whenever the operation conditions of the transmitter change, e.g., carrier frequency, bandwidth, power and potentially the beam direction. We consider the number of floating point operations per sample (FLOPs) as the complexity metric. For clarity, the notations used in this section are summarized as follows: $\ceil{.}$ denotes the ceil operator, $N_{\text{IBF}}$ stands for the total number of instantaneous BFs, i.e., the filter lags are excluded from this number, while $N_{\text{BF}}=N_{\text{IBF}}D_\text{DPD}$ denotes the total number of BFs, where $D_{\text{DPD}}$ refers to the number of taps of an individual DPD filter. Additionally, $D_{\text{lin}}$ denotes the length of the linear filter(s) used for linear response modeling in the adopted CL DPD in order to generate the error signal. Furthermore, as earlier, $L$ is the number of subarrays, $N_{\text{CL}}$ and $N_{\text{ILA}}$ denote the learning block-sizes of the CL DPD and the ILA-based DPD systems, respectively, while $I_{\text{CL}}$ and $I_{\text{ILA}}$ are the corresponding numbers of block iterations for the two techniques. Lastly, $T$ is a variable that reads $T = L$ and $T= 1$ when analyzing the CL DPD complexity with and without crosstalk, respectively. This takes into account the difference in the error signal generation complexity in cases without and with crosstalk.

The complexity expressions in their general form are gathered in Table \ref{tab:complexity}, while the exact complexity numbers for a specific evaluation scenario are provided along with the numerical results in Section \ref{Results}. To derive the expressions in Table \ref{tab:complexity}, a complex multiplication is assumed to cost 6 FLOPs.}

{It can be generally noted that inverting strongly nonlinear systems commonly calls for relatively high polynomial orders, especially if very high-quality linearization is pursued. In mmWave active arrays, when emphasizing power-efficiency, the individual PA units are operating in large compression close to saturation \cite{GreenComm,Swedes_review}, and thus the system can be considered strongly nonlinear implying that the level of nonlinear distortion is high. While in our following numerical experiments we utilize memory polynomials of order 11, to demonstrate very high-quality linearization of measured mmWave PA units, it is noted that in practice sufficient linearization may also be obtained with lower polynomial orders, particularly since the linearity requirements at mmWaves \cite{3GPPTS38104} are somewhat relaxed compared to lower-frequency systems. }

\section{Numerical Results}\label{Results}
In this section, a quantitative analysis of the performance of the considered CL DPD architecture and parameter learning solution is pursued and presented by means of comprehensive Matlab simulations.  

\subsection{Evaluation Environment and Assumptions}
The evaluation environment builds on the clustered mmWave channel model described in Subsection II-B, containing $C=6$ clusters each with $R=5$ rays. We assume that a LOS component is always available and that the Ricean K-factor is 10~dB \cite{3GPPTR38901}. The maximum considered excess delay is 60 ns, a number that is well inline with the assumptions in \cite{3GPPTR38901}. We further assume that a hybrid MIMO transmitter simultaneously serves $U=2$ single-antenna users. {The overall transmitter is assumed to contain $L=2$ TX chains and subarrays, each of them having $M=32$ antenna elements and the corresponding PA units. Therefore, a total of $64$ antennas and PAs are considered. In each subarray, the antenna spacing is half the wavelength. It is noted that the developed methods are not limited to any particular values of $U$, $L$ or $M$, while the above values reflect an implementation feasible example scenario. Additionally, it is noted that no simplifying assumptions or approximations are taken in the numerical experiments but the received signals are numerically calculated in their exact forms.}

\begin{figure}[t!]
\centering
\includegraphics[width=.75\linewidth]{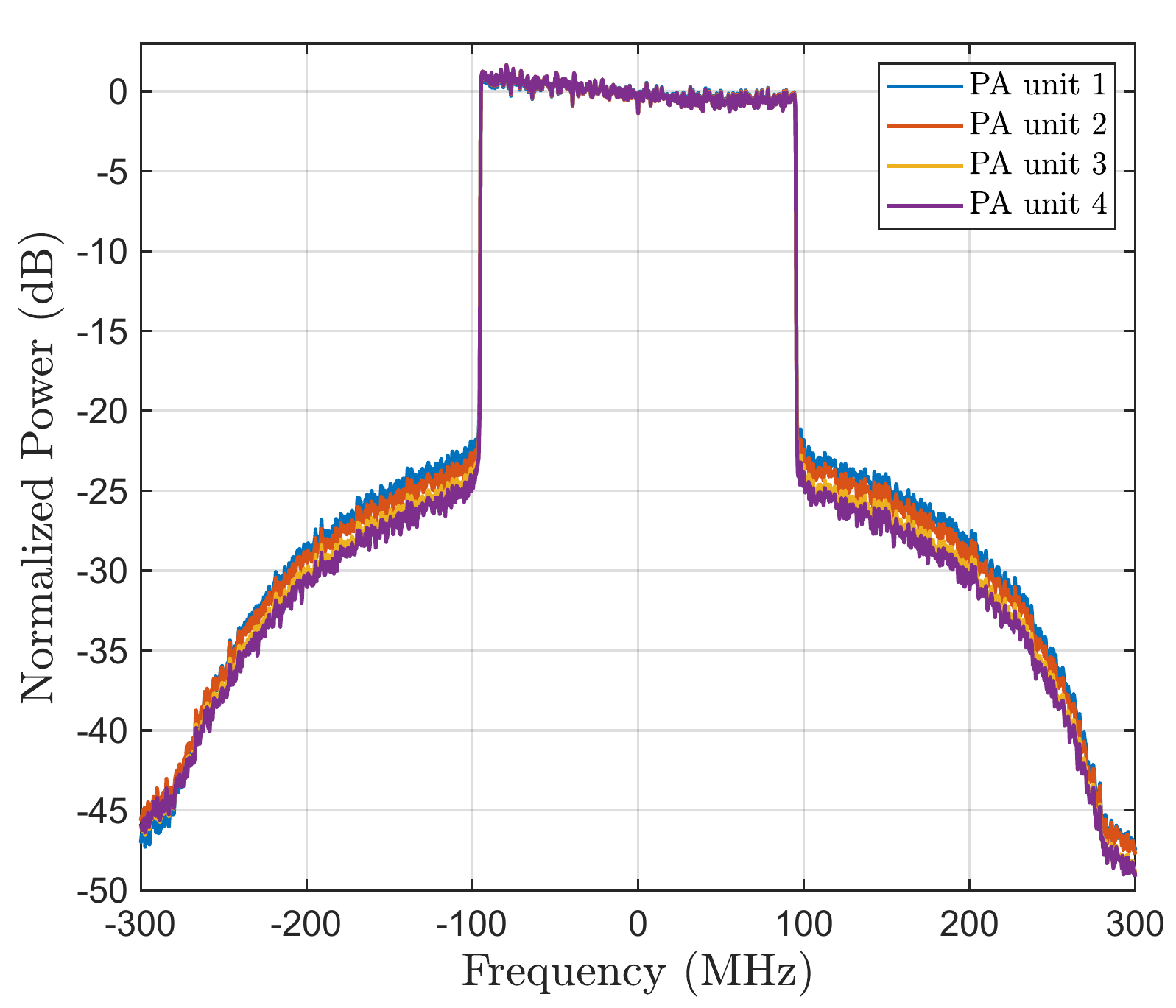}
\caption[]{Example normalized individual PA output spectra of $4$ different PA models extracted from measurements on a set of mmWave PAs operating at 28 GHz carrier frequency.}
\label{Fig.PA_outputs}
\end{figure}

{Furthermore, we evaluate the performance of the adopted DPD solution for both the single-beam and multi-beam analog beamformers}, discussed in Section \ref{Design_Precoders}, though we put most emphasis on multi-beam case as that reflects the true multiuser processing more genuinely. To design the analog beamforming phases, we use the optimization approach described in \cite{AnalogBeamforming}, with angle information as the input. Subcarrier-wise digital precoders are always calculated through the ZF approach, as shown in (3), complemented with proper sum-power normalization. In the basic evaluations, perfect channel state information (CSI) is assumed to be available at the transmitter, in terms of the angles as well as equivalent subcarrier-level complex channel matrices, while the impact of imperfect CSI is addressed in Sub-section \ref{sec:imperfect_CSI}. 200 MHz carrier bandwidth is assumed as a representative number in mmWave systems, conforming to 3GPP 5G NR specifications \cite{3GPPTS38104} with OFDM subcarrier spacing of 60 kHz, $K_{\mathrm{ACT}} = 3168$ active subcarriers and FFT size of $K_{\mathrm{FFT}}=4096$. Finally, time-domain windowing is adopted to improve the spectral containment of the OFDM signals, while the peak-to-average-power ratio (PAPR) of the composite multicarrier waveform in each TX chain is limited to $8.3$~dB, through iterative clipping and filtering (ICF). {No beam-tapering is applied in these results, but it is noted that the methods work also with tapering.}

{For modeling the individual PA units, measurement data from an actual set of mmWave PAs, {specifically the PA units in Anokiwave AWMF-0108 ICs,} operating at 28.0~GHz carrier frequency is used, and memory polynomials of order $P=11$ and memory depth $D_{\text{DPD}}=3$ are identified utilizing the same test signal as the one described above. The measured PA models are also shared along the article, {together with other sets of PA models obtained by measuring Analog Devices HMC943APM5E PA ICs at 28.0~GHz and 28.5~GHz. The modeling accuracy of the measured memory polynomial models is in the order of 33~dB in terms of normalized mean squared error (NMSE).}} Example representative power spectra of four PA output signals are shown in Fig. \ref{Fig.PA_outputs}, where substantial nonlinear distortion as well as mutual differences between the characteristics of the individual PA units can be observed. 

\begin{figure}[t!]
\centering
\includegraphics[width=.75\linewidth]{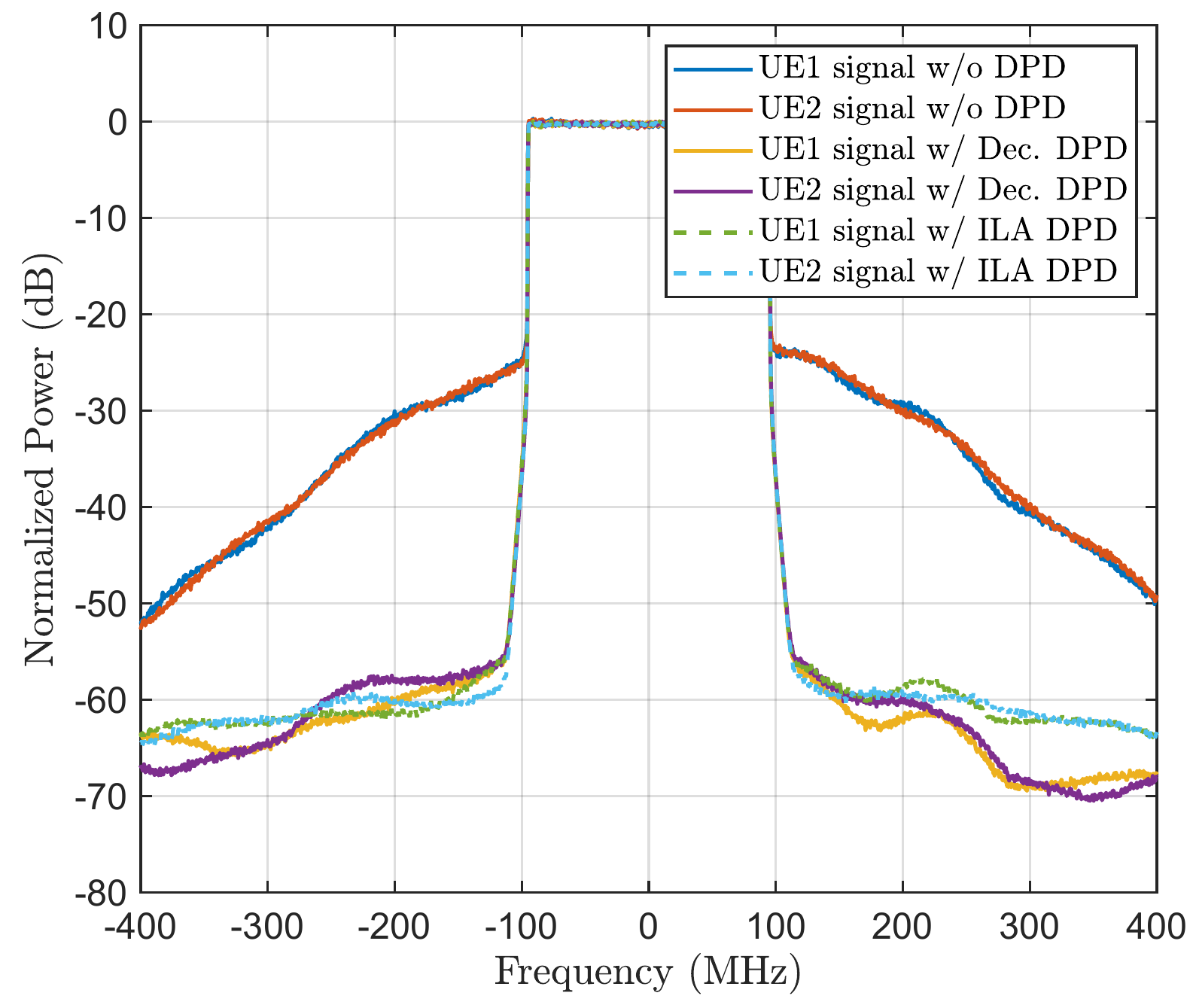}
\caption[]{Normalized combined spectra at the two intended users, without and with DPD, when the \textit{multi-beam analog beamformer} is adopted. For reference and comparison purposes, both the CL DPD and the reference ILA DPD results are shown. No crosstalk is assumed.}
\label{RX_spectrumsUE}
\end{figure}

As the basic performance metrics, we consider the error vector magnitude (EVM) and adjacent channel leakage ratio (ACLR) to evaluate the inband signal quality as well as the corresponding adjacent channel interference due to spectrum regrowth, respectively, as defined in \cite{3GPPTS38104} and \cite{3GPP_BS}, and both interpreted for the combined signals. 
The EVM is defined as
\begin{equation}
\text{EVM}_{\%} = \sqrt{P_{\text{error}}/P_{\text{ref}}} \times 100\%,
\end{equation}
where $P_{\text{error}}$ is the power of the error between the ideal symbols and the corresponding symbol rate complex samples of the combined array output at the intended receiver direction, {after amplitude and phase equalization at subcarrier level}, with both signals normalized to the same average power, while $P_{\text{ref}}$ is the reference power of the ideal signal. {It is noted that the ICF-based PAPR mitigation processing results in an EVM floor of ca. $2\%$ which thus serves as the reference in the evaluations}. On the other hand, the ACLR is defined as the ratio between the power observed at the intended channel, $\text{P}_{\text{intended}}$, and that at the right or left adjacent channels, $\text{P}_{\text{adjacent}}$, interpreted for the effective
combined signal at the RX, and expressed as
\begin{equation}
\text{ACLR}_{\text{dB}} = 10 \log_{10} \frac{\text{P}_{\text{intended}}}{\text{P}_{\text{adjacent}}}.
\end{equation}
{It is noted that the ACLR levels in mmWave systems, without linearization, are commonly very low \cite{3GPPTS38104}, in the order of only 25~dB, reflecting thus very nonlinear system and PA saturation.}

In all the following numerical results, the DPD nonlinearity order $Q=11$ and the memory depth $D_{\text{DPD}}=3$ in both ($L=2$) DPD units, which results in $N_{\text{IBF}} = 6$ and $N_{\text{BF}} = 18$. The parameter estimation is carried out with the decorrelation-based approach, implemented in a block-adaptive manner, such that each block contains $N_{\text{CL}} = 20,000$ samples and a total of $I_{\text{CL}}=15$ iterations are used. Thus, overall, the DPD parameter estimation utilizes 300,000 complex samples. Furthermore, for the error signal generation in parameter learning, {the effective linear loop and coupling responses, defined along (\ref{error_signal}) or (\ref{eq:z_fb_crosstalk}), are estimated through ordinary block least-squares, with 20,000 samples, prior to the DPD parameter learning.} 
{For reference and comparison purposes, also single-input ILA DPD results are commonly shown, building on least-squares parameter learning. The same amount of learning samples as in CL DPD is utilized, in the form of $I_{\text{ILA}}=3$ ILA iterations each containing $N_{\text{ILA}} = 100,000$ samples.}
\begin{table}[t!]
\setlength{\tabcolsep}{2pt}
\renewcommand{\arraystretch}{1.2}
\caption{EVM and ACLR results without crosstalk and when multibeam analog beamforming is adopted}
\centering
{\begin{tabular}{lll}\hline
                                  &\:\:\: \textbf{EVM ($\%$) } &\:\:\: \textbf{ACLR L\:/\:R (dB)}    \\\hline
Without DPD at UE1                      &\:\:\: 6.41        &\:\:\: 27.19\:/\:28.94       \\
Without DPD at UE2 				  &\:\:\: 7.42       &\:\:\: 27.03\:/\:28.73       \\
With Closed-loop DPD at UE1							  &\:\:\: 2.25        &\:\:\: 60.41\:/\:59.73 \\   
With Closed-loop DPD at UE2							  &\:\:\: 2.32        &\:\:\: 59.45\:/\:58.41\\
With ILA DPD at UE1							  &\:\:\: 2.31        &\:\:\: 59.45\:/\:58.41\\
With ILA DPD at UE2							  &\:\:\: 2.35        &\:\:\: 59.61\:/\:58.78\\\hline
\end{tabular}}{}
\label{tab:EVM_ACLR}
\end{table}

\begin{table}[t!]
\centering
\setlength{\tabcolsep}{2pt}
\renewcommand{\arraystretch}{1.3}
\caption{\textsc{{DPD main path processing complexity per sample and total DPD learning complexity of the hybrid MIMO transmitter for a total of 300,000 samples}}}
\label{tab:complexity_num}
\begin{tabular}{cccc}
\cline{1-4}

 & &\textbf{CL DPD (w/o ; w/ crosstalk)}  & \textbf{ILA DPD} \\ \hline
\multicolumn{1}{c}{\multirow{3}{*}{\makecell{\textbf{DPD} \\ (FLOPs/sample)}}} & BF gen. & 22  & 22  \\ 
\multicolumn{1}{c}{} & Filtering& 272 & 284 \\ 
\multicolumn{1}{c}{} & Total & 294 & 306  \\ 

\multicolumn{1}{c}{\multirow{3}{*}{\makecell{\textbf{Learning} \\ (MFLOPs)}}} & BF gen. & $-$ & 7.8  \\  
\multicolumn{1}{c}{} & DPD est. & 102.31 ; 116.95& 777.64  \\ 
\multicolumn{1}{c}{} & Total & 102.31 ; 116.95 & 785.44  \\ \cline{1-4} 
\end{tabular}
\end{table}
\subsection{Results without Crosstalk}
\subsubsection{DPD Performance at Intended Receivers}
First, we evaluate and demonstrate the performance of the CL DPD structure and parameter learning solution from the two intended receiver directions point of view. The 64 PA output signals combine through their respective frequency-selective channels towards the intended receivers, and the corresponding power spectra of the effective combined signals are depicted in Fig. \ref{RX_spectrumsUE}, without and with DPD. Furthermore, the \textit{multi-beam analog beamformer} approach is considered in this example figure, and therefore both subarrays provide simultaneous beams towards both users. Very similar combined signal spectra are obtained when the \textit{single-beam analog beamformer} is adopted, and are thus not explicitly shown. Table \ref{tab:EVM_ACLR} shows the corresponding numerical EVM and ACLR values, demonstrating excellent linearization performance at both intended users. {It can also be observed that very similar linearization performance can be obtained through both the CL DPD and the reference ILA DPD, as the crosstalk aspects are not yet considered. The computational complexities associated to both DPD solutions are shown in Table~\ref{tab:complexity_num}, evidencing the low complexity stemming from the described CL DPD compared to the widely adopted ILA DPD, which is essential towards real-time implementation and tracking of fast changes in the operation conditions of the array. {It is also noted that with lower DPD polynomial orders, the complexity would further reduce, though naturally the linearity would also degrade.}}

\begin{figure}[t!]
\centering
\includegraphics[width=.75\linewidth]{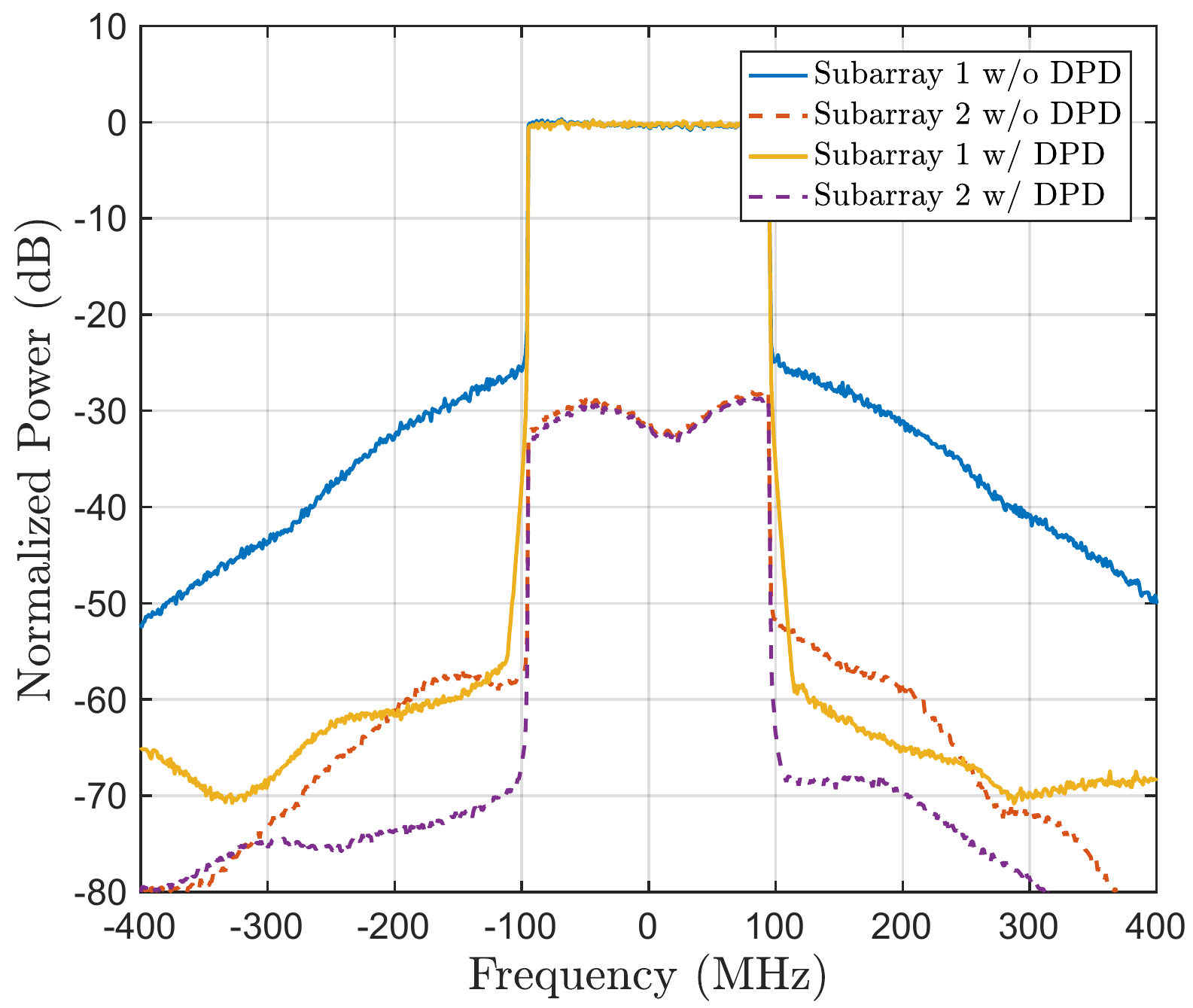}
\caption[]{Normalized spectra of the received combined signals at UE 1, stemming from individual transmit subarrays, considering the \textit{single-beam analog beamformers}. Total received signal is not shown. 'With DPD' refers to the CL DPD method. No crosstalk is assumed. 
}
\label{Fig.RF_1_Individuals}
\end{figure}

\begin{figure}[t!]
\centering
\includegraphics[width=.75\linewidth]{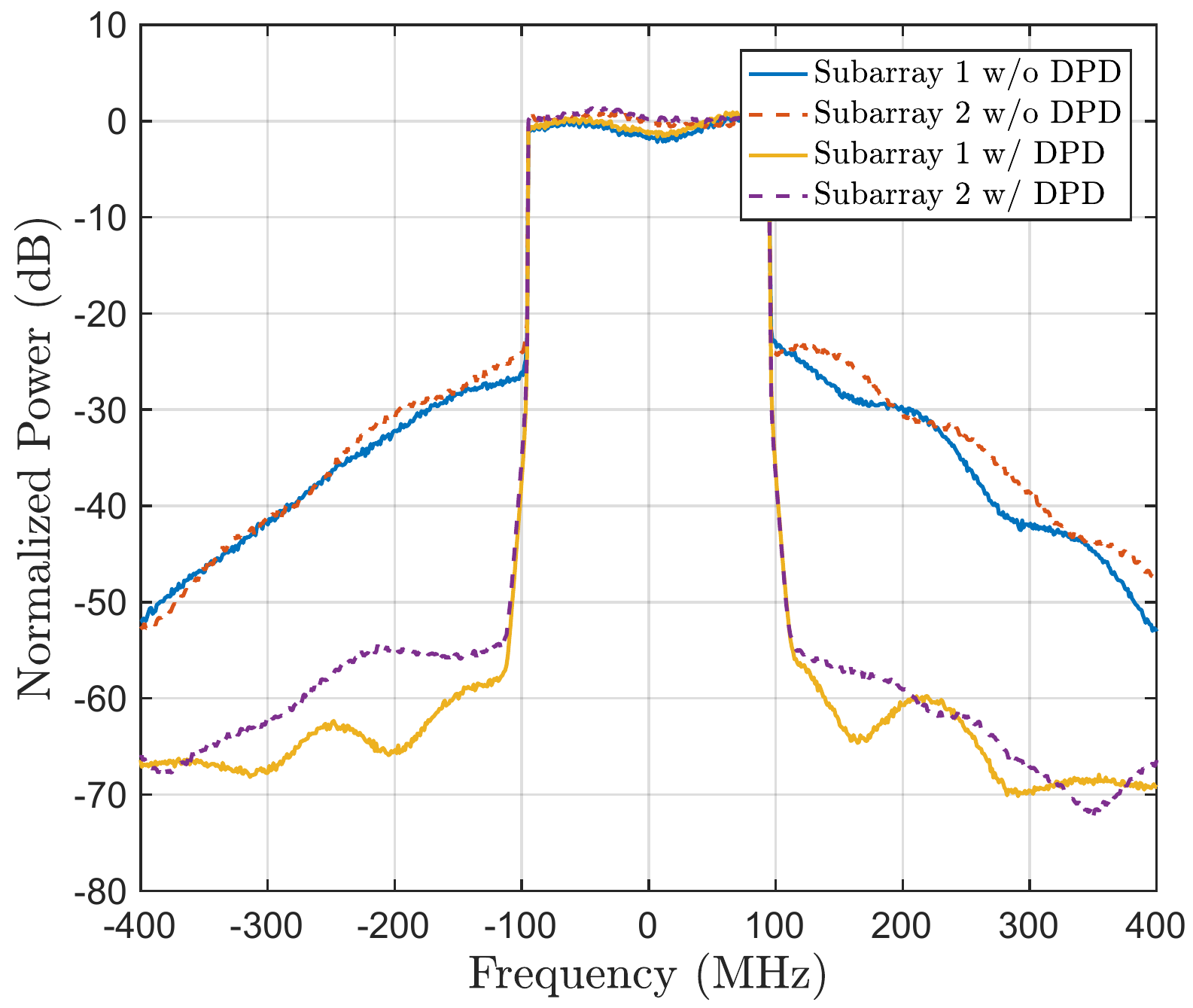}
\caption[]{Normalized spectra of the received combined signals at UE 1, stemming from individual transmit subarrays, considering the \textit{multi-beam analog beamformers}. Total received signal is not shown. 'With DPD' refers to the CL DPD method. No crosstalk is assumed.}
\label{Fig.RF_2_Individuals}
\end{figure}

Despite the total combined signal qualities at the intended receivers are very similar for both single-beam and multi-beam analog beamformers, there are fundamental differences in how the DPD processing contributes to suppressing the combined nonlinear distortion in these two cases. To explore this further, we next illustrate the combined received signal spectra at one of the intended users, say UE 1, and deliberately consider the contributions of the two TX subarrays separately.
First, when the \textit{single-beam analog beamformer} is considered, the spectra of the combined subarray signals are shown in Fig. \ref{Fig.RF_1_Individuals}, without and with DPD. Now, due to the single-beam analog beamformer, the received signal at UE 1 is largely dominated by subarray 1 while the contribution of subarray 2 is substantially weaker. Hence, as can be observed in the figure, the linearization impact of the DPD unit of subarray 1 is substantial, while it is the combined effect of the array isolation and DPD processing that reduces the OOB emissions stemming from subarray 2. The behaviors of the combined subarray signal spectra at UE 2 are very similar, with the roles of the subarrays interchanged, and are thus omitted.

\begin{figure}[t!]
\centering
\includegraphics[width=.75\columnwidth]{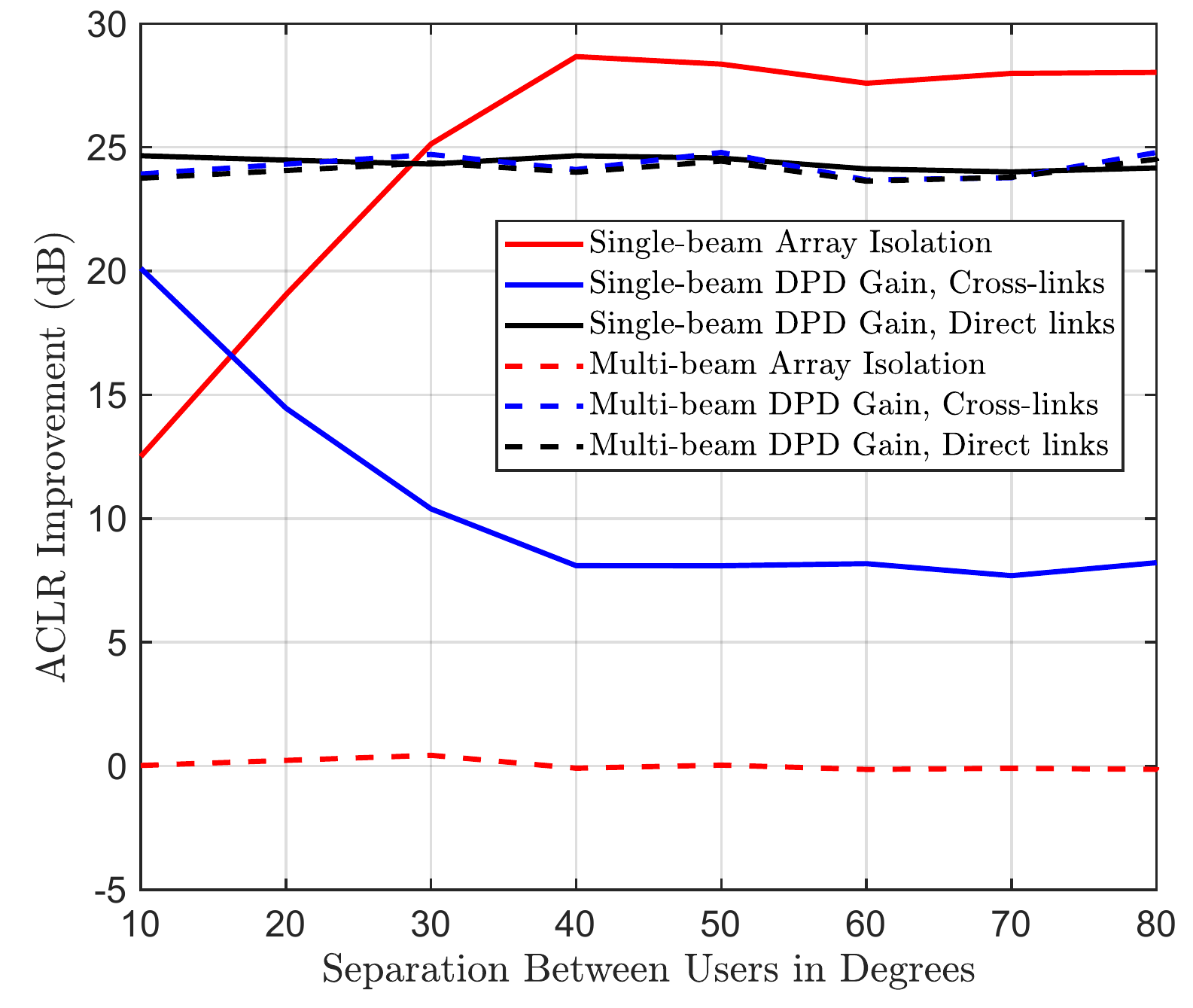}
\caption[]{Impact of the array isolation and the DPD processing on the combined OOB power when the \textit{single-beam analog beamformers} (solid curves) and \textit{multi-beam analog beamformers} (dashed curves) are considered. No crosstalk is assumed. 
}
\label{Fig.Array_DPD_gain}
\end{figure}

On the other hand, when the \textit{multi-beam analog beamformer} is adopted, there is then coherent combining taking place from both subarrays towards the considered UE 1. In this case, the array isolation does not essentially help in controlling the OOB emissions but as shown in Fig. \ref{Fig.RF_2_Individuals}, the CL DPD units can now simultaneously linearize the combined signals of multiple beams. Therefore, the good OOB reduction is primarily due to the DPD units. Again, the received spectra at the UE 2 behave very similarly, and are thus omitted.

\subsubsection{Impact of DPD gain and array isolation} To provide further insight on the roles of the array isolation and the DPD, we continue to explore the two-user scenario such that the angular separation between the two users is varied. We first place the two intended users very close to each other in the angular domain and configure the analog beams accordingly. Their channel responses are thus very similar, except for the exact phase differences due to the geometry of the environment and scattering. Under these assumptions, highly coherent propagation is expected from both subarrays towards the two intended users regardless of the chosen RF beamforming strategy. Then, the location of one of the intended receivers is kept fixed,  while the other one gradually moves along a circular trajectory such that the angular separation is increasing, and beamformers are always adjusted accordingly, while the DPD coefficients are kept fixed for all directions.

The obtained results in terms of the relative ACLR behavior can be found in Fig. \ref{Fig.Array_DPD_gain}  when the single-beam  and the multi-beam analog beamformers are adopted, averaged over 100 independent channel realizations for each angular separation value. In the figure, we show separately the behavior of the combined out-of-band emissions due to the two subarrays for the so-called direct links (subarray 1 to UE 1 and subarray 2 to UE 2, averaged across the two users) and the so-called cross-links (subarray 1 to UE 2 and subarray 2 to UE 1, averaged again across the two users). The \textit{Array Isolation} refers to the ratio of the combined OOB emissions of the direct links and those of the crosslinks, when the DPD processing units are deliberately set off. The \textit{DPD Gain}, in turn, refers to the average ACLR improvement obtained by using the CL DPD units, evaluated separately for the cross-links and the direct links.

In the single-beam beamformer case, when the users are close in angular domain, the array isolation is naturally small while the DPDs provide good linearization also for the cross-links, both aspects being due to the very high similarity between the array channels of the direct and cross-links. On the other hand, as the angular separation starts to increase, the DPD performance at the cross-links decays while the array isolation increases, but the corresponding total gain stays essentially constant. Then, when the multi-beam analog beamformers are adopted, both users essentially experience coherent propagation from both subarrays. In this case, as expected, the array gain is essentially zero while large DPD gains are systematically available for both the direct and the cross-links independent of the angular separation.

These results show that in the case of \textit{multi-beam analog beamformer}, the DPD units provide simultaneous linearization from each subarray towards all users. Additionally, when the \textit{single-beam analog beamformers} are adopted, the combined effect of array isolation and DPD processing will keep the combined OOB power low. Overall, the results and findings along Figs. 4-7 confirm many of the basic hypotheses made in the previous technical sections. Specifically, the results demonstrate and verify that a single DPD unit can linearize a bank of different PAs when viewed from the combined signal point of view. Additionally, the results verify that the DPD units can provide linearization simultaneously towards multiple directions at which coherent combining is taking place, i.e., when multi-beam analog beamformers are adopted.

\subsubsection{DPD Performance in Spatial Domain at Intended and Victim Users}\label{sub:spatial_analysis}

While the above examples demonstrate very high-quality linearization at intended receivers in snap-shot like scenarios, we next pursue evaluating the behavior of the unwanted emissions in the overall spatial domain, i.e., at randomly placed intended and victim users. {In case of the victim receivers, the ACLR is calculated as the ratio between the power radiated at the intended channel in the direction of the intended user, and the OOB power radiated at the adjacent channel in the direction of the victim receiver. Additionally, the OOB power is always measured from the particular neighboring channel -- left or righ -- whichever is worse, such that the most challenging cases are reflected in the analysis.}
In these evaluations, we first drop the two intended users at randomly drawn directions and calculate the analog and digital beamformers accordingly. {\color{black}In analog domain, multi-beam approach is utilized.} The DPD system parameter learning is then executed, and after that, while keeping the beamformer and DPD coefficients fixed, we drop 10,000 victim receivers
at randomly drawn directions, and evaluate the OOB emissions at all these victim receivers. This is then further iterated over different randomly drawn intended RX directions, such that the beamformer coefficients are recalculated, while experimenting two different scenarios of re-executing and not re-executing the DPD parameter learning. 
Finally, empirical distributions of the ACLRs at the victim receivers as well as at the intended receivers are evaluated.

\begin{figure}[t!]
\centering
\subfigure{
\includegraphics[width=.75\linewidth]{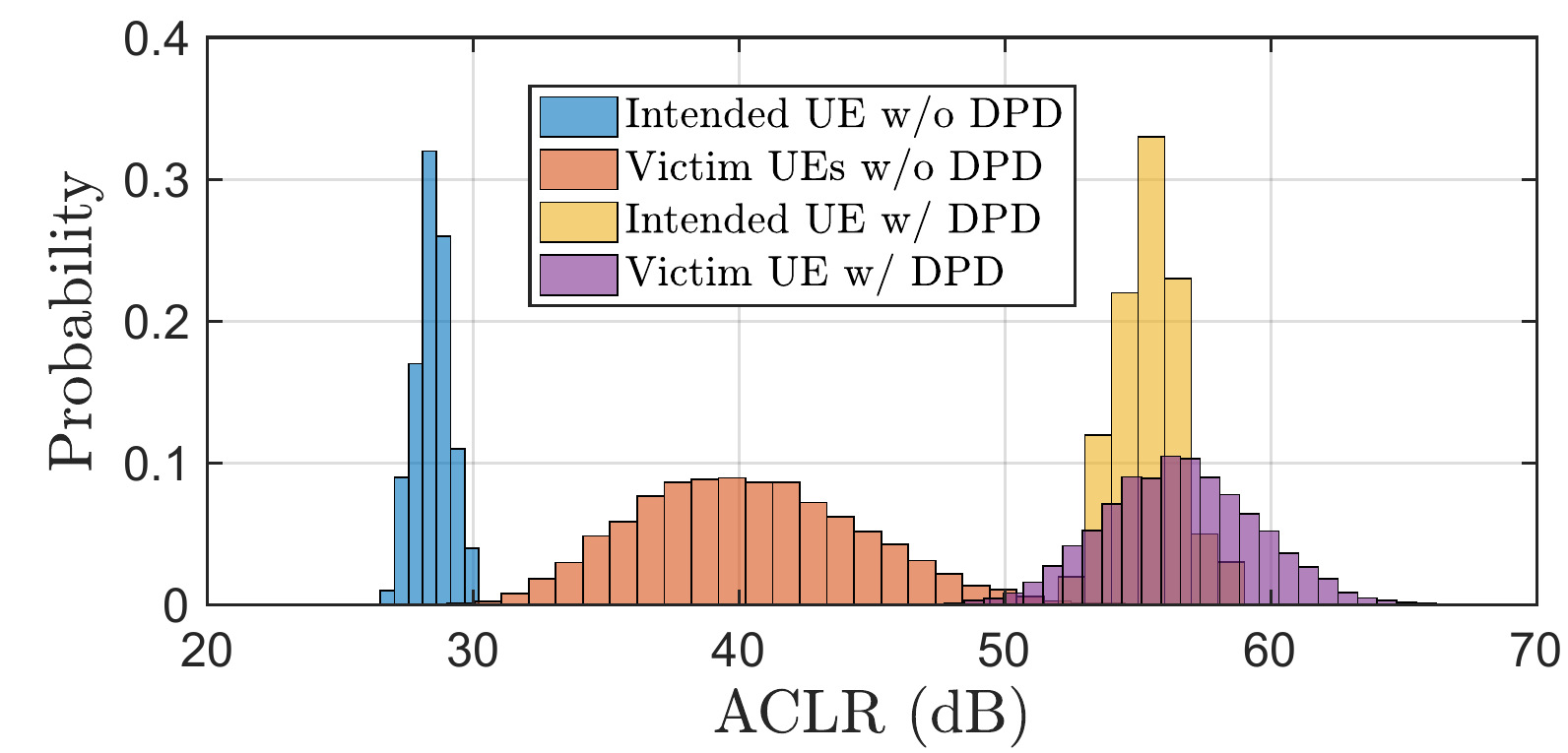}
}
\subfigure{
\includegraphics[width=.75\linewidth]{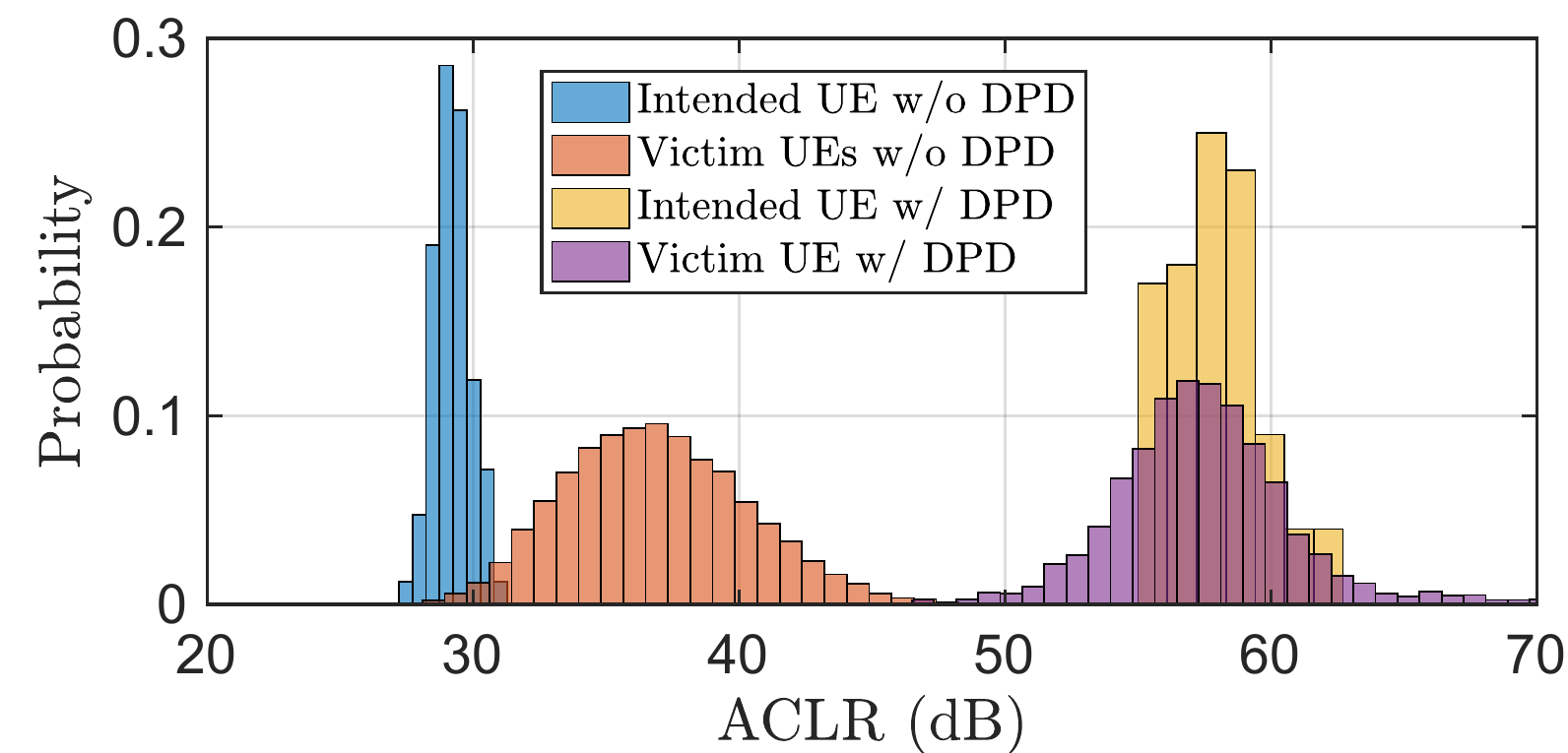}
}
\caption{{Empirical ACLR distributions at intended and victim UEs  (\emph{i}) when the DPD parameter learning is carried out separately for each random drop of intended UEs (top) and (\emph{ii}) when the DPD parameter learning is executed only for the first drop while fixed for others (bottom). 'With DPD' refers to the closed-loop DPD method. No crosstalk is assumed.}}
\label{fig:spatial_domain_nocrosstalk}
\end{figure}

The obtained empirical ACLR distributions are shown in Fig. \ref{fig:spatial_domain_nocrosstalk}, {covering the two scenarios in terms of DPD learning. In the first scenario, illustrated on top, the DPD learning is re-executed for every new drop of the intended UE locations, while in the second scenario, illustrated in bottom, the DPD learning is executed only for the first drop of the intended UEs while kept fixed afterwards}. 
Firstly, the ACLR distributions at intended UEs show that the DPD units systematically and reliably improve the signal quality, independent of whether new DPD parameter learning is executed for new array channel realizations or not. Additionally, 
the ACLR distributions at victim receivers without any DPD processing clearly indicate that the exact ACLR can vary relatively widely depending on the exact array channel realizations. However, when the DPD units are turned on, large systematic ACLR improvement is obtained with the minimum ACLR realization being ca. 50~dB. Overall, these results show that systematic and reliable linearization can be provided, at both intended and victim receivers, through the adopted approach, and also that changes in the array channels do not call for new DPD parameter learning. As shown in the next subsection, this latter conclusion is, however, valid only when there is no crosstalk.

{\subsection{Results with Crosstalk}
We next experiment the nonlinear distortion behavior and the CL DPD system linearization performance under crosstalk. As representative  baseline values, we set the PA input crosstalk level, within an individual subarray, between two neighboring branches to $-$20~dB while the corresponding antenna crosstalk level is set to $-$10~dB \cite{MIMO_DPD_2,Prediction_distortion,28GHz_crosstalk}. The other coupling factors between any other two PA input or antenna branches are then assumed to decay as a function of the square of the physical distance and are calculated accordingly. The exact phases of the crosstalk terms, in turn, are always calculated through the effective physical distance between the coupling points, relative to the wavelength. The baseline strength of the crosstalk between two neighboring antenna units is also varied in the evaluations. 

An example spectral illustration, under crosstalk and assuming the multi-beam analog beamformer case, is given in Fig. \ref{RX_spectrumsUE_crosstalk}, covering also the single-input ILA DPD for reference. As can be observed, the ILA-based reference method is clearly compromised by the crosstalk, while the CL method provides excellent linearization performance despite the relatively strong crosstalk. The corresponding exact EVM and ACLR numbers are shown in Table \ref{tab:EVM_ACLR2}, while the corresponding processing complexities are shown in Table \ref{tab:complexity_num}, again evidencing the lower complexity of the CL DPD solution compared to ILA-based DPD.

\begin{figure}[t!]
\centering
\includegraphics[width=.75\linewidth]{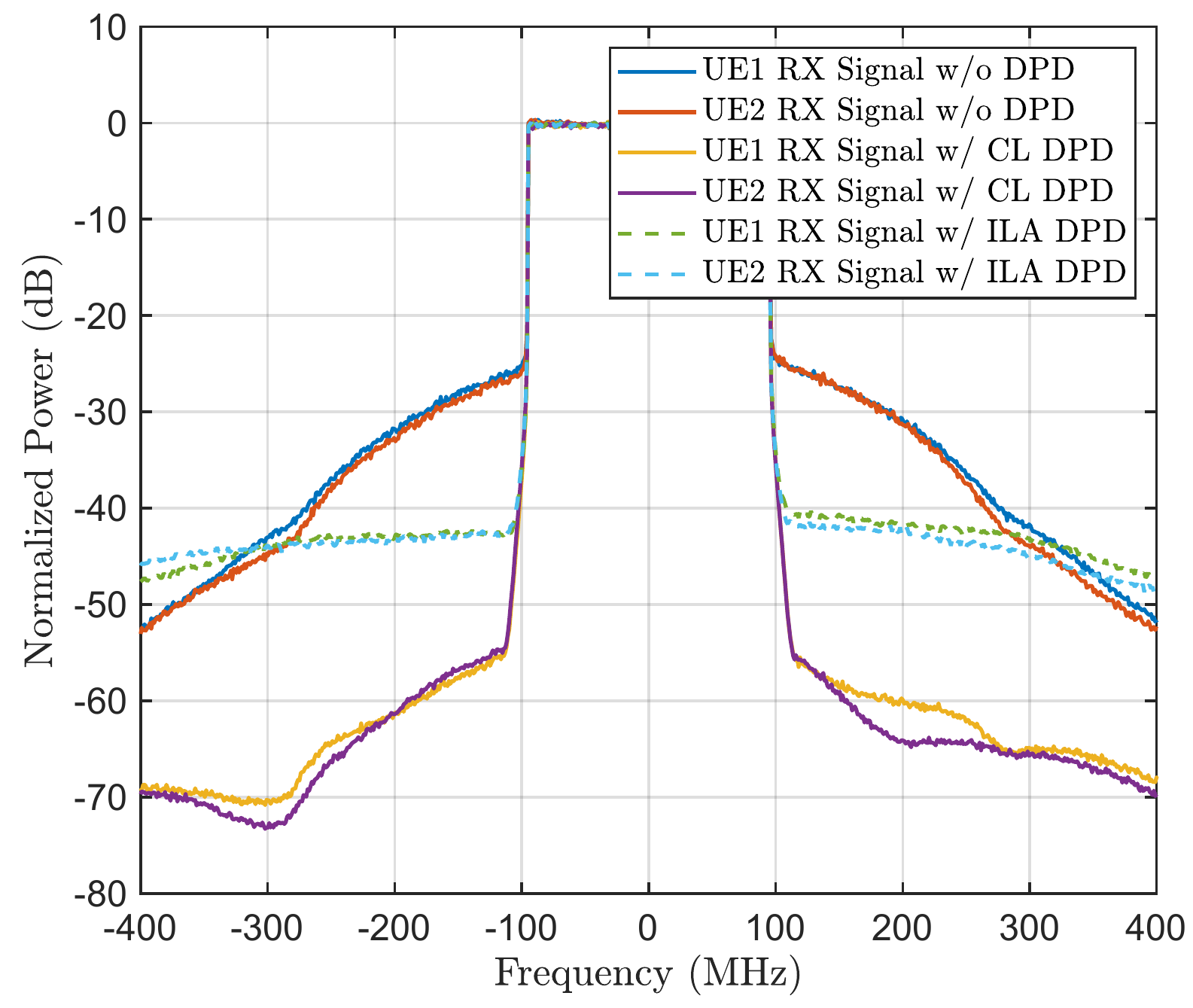}
\caption[]{{Normalized combined spectra at the two intended users, without and with DPD, when the \textit{multi-beam analog beamformers} are adopted and $M=32$. Baseline PA input crosstalk level is $-$20~dB while the baseline antenna crosstalk level is $-$10~dB.}}
\label{RX_spectrumsUE_crosstalk}
\end{figure}

\begin{table}[t!]
\setlength{\tabcolsep}{2pt}
\renewcommand{\arraystretch}{1.2}
\caption{EVM and ACLR results with crosstalk and when multibeam analog beamforming is adopted}
\centering
{\begin{tabular}{lll}\hline
                                  &\:\:\: \textbf{EVM ($\%$)}  &\:\:\: \textbf{ACLR L\:/\:R (dB)}    \\\hline
Without DPD at UE1                      &\:\:\: 7.91        &\:\:\: 29.69\:/\:29.92      \\
Without DPD at UE2 				  &\:\:\: 7.42       &\:\:\: 29.47\:/\:29.90       \\
With Closed-loop DPD at UE1							  &\:\:\: 2.89        &\:\:\: 58.99\:/\:58.66 \\ 
With Closed-loop DPD at UE2							  &\:\:\: 3.09        &\:\:\: 57.32\:/\:58.21\\
With ILA DPD at UE1							  &\:\:\: 4.61        &\:\:\: 40.99\:/\:41.12\\
With ILA DPD at UE2							  &\:\:\: 4.55        &\:\:\: 41.67\:/\:41.39.\\\hline
\end{tabular}}{}
\label{tab:EVM_ACLR2}
\end{table}

To further experiment the impact of the crosstalk on the linearization performance, we next vary the antenna crosstalk level from $-$15~dB to $-$10~dB, and evaluate the  ACLR performance for both ILA and CL DPD methods. The results are shown in Fig. \ref{fig:aclr_vs_crosstalk}, clearly evidencing the superiority and excellent robustness of the CL DPD under different crosstalk levels. The figure is also showing the corresponding ACLR performance when the subarray size is reduced to $M=8$.} {In this case, the superiority of the CL DPD method is even more obvious, compared to ILA-based DPD, since now there is less help from the noncoherent combining of the crosstalk terms in the combined feedback signal, compared to the case of $M=32$.}

{Next, we pursue spatial domain analysis similar to the one in Fig. \ref{fig:spatial_domain_nocrosstalk}, but now under the baseline crosstalk levels of $-$20~dB and $-$10~dB. The obtained ACLR distributions are shown in Fig.~\ref{fig:spatial_domain_crosstalk}. As can be observed, when the DPD learning is re-executed for every new intended UE drop (new array channel realization for intended UEs, and hence new analog and digital beamforming coefficients), the adopted CL DPD system provides excellent linearization comparable to the crosstalk-free case in Fig. \ref{fig:spatial_domain_nocrosstalk}. However, if the DPD coefficients are kept fixed, while the beamforming still follows the changes in the involved channels, a systematic loss in the linearization performance can be observed. This is stemming from the beam-dependency of the nonlinear distortion, caused by the crosstalk \cite{Swedes_review,Low_number_steering}. These findings thus imply that continuous or close-to continuous DPD learning is needed, which in turn raises the importance of the computational simplicity of the CL single-input DPD unit and the involved decorrelation based learning. Additionally, the combined feedback observation receiver facilitates the tracking of the changes on the operation conditions due to beamforming. {Alternatively, one could consider a predefined set of DPD coefficients for different beam directions \cite{Low_number_steering}. However, adaptability to changes due to, e.g., temperature drifts or device aging would be limited.}}

\begin{figure}[t!]
\centering
\includegraphics[width=.75\linewidth]{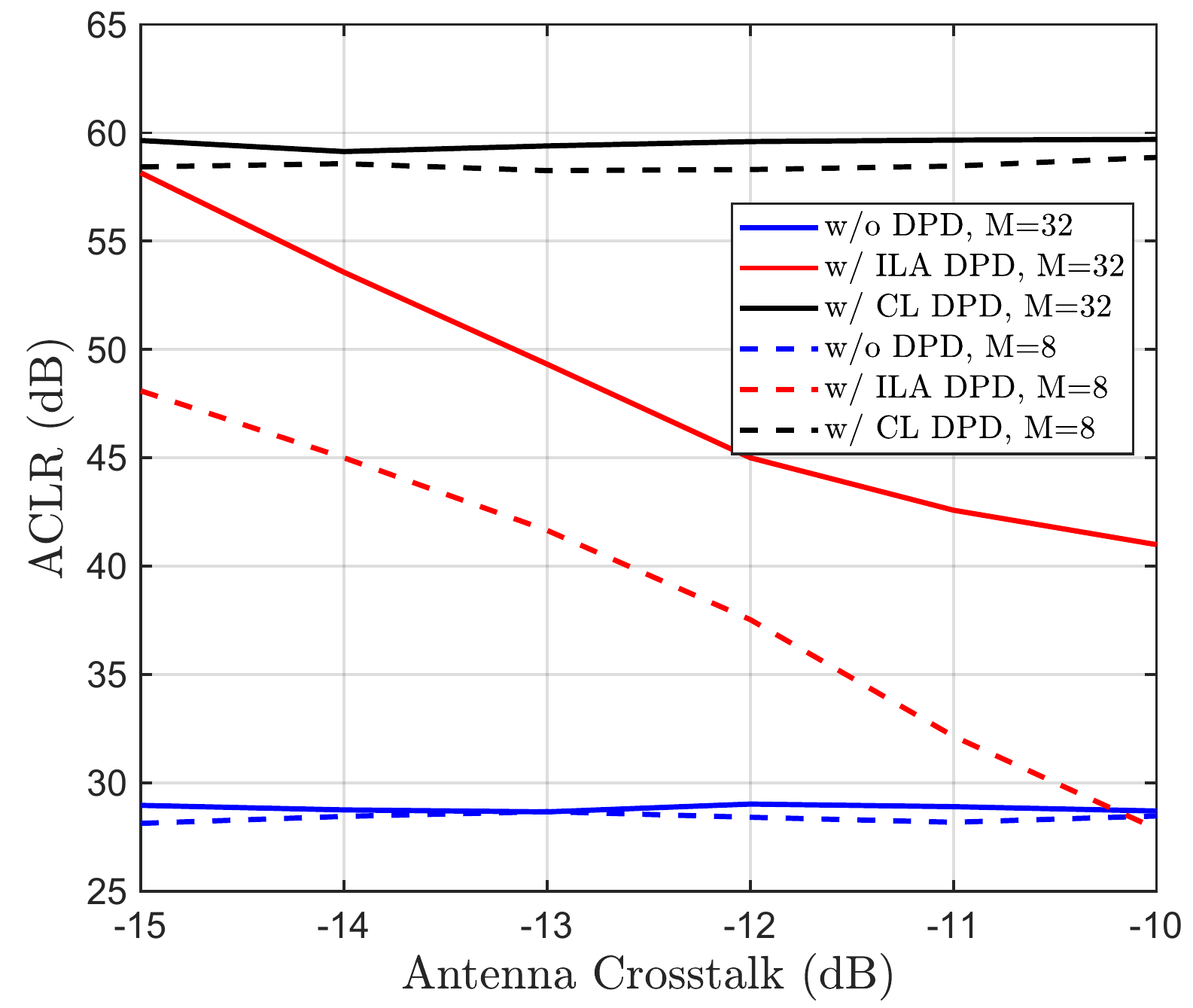}
\caption[]{{Obtained ACLR performance as a function of the baseline antenna crosstalk level. The solid lides correspond to the basic evaluation settings (with $M=32$) while the dashed curves correspond to case with reduced subarray size of $M=8$.}}\label{fig:aclr_vs_crosstalk}
\end{figure}
\begin{figure}[t!]
\centering
\subfigure{
\includegraphics[width=.75\linewidth]{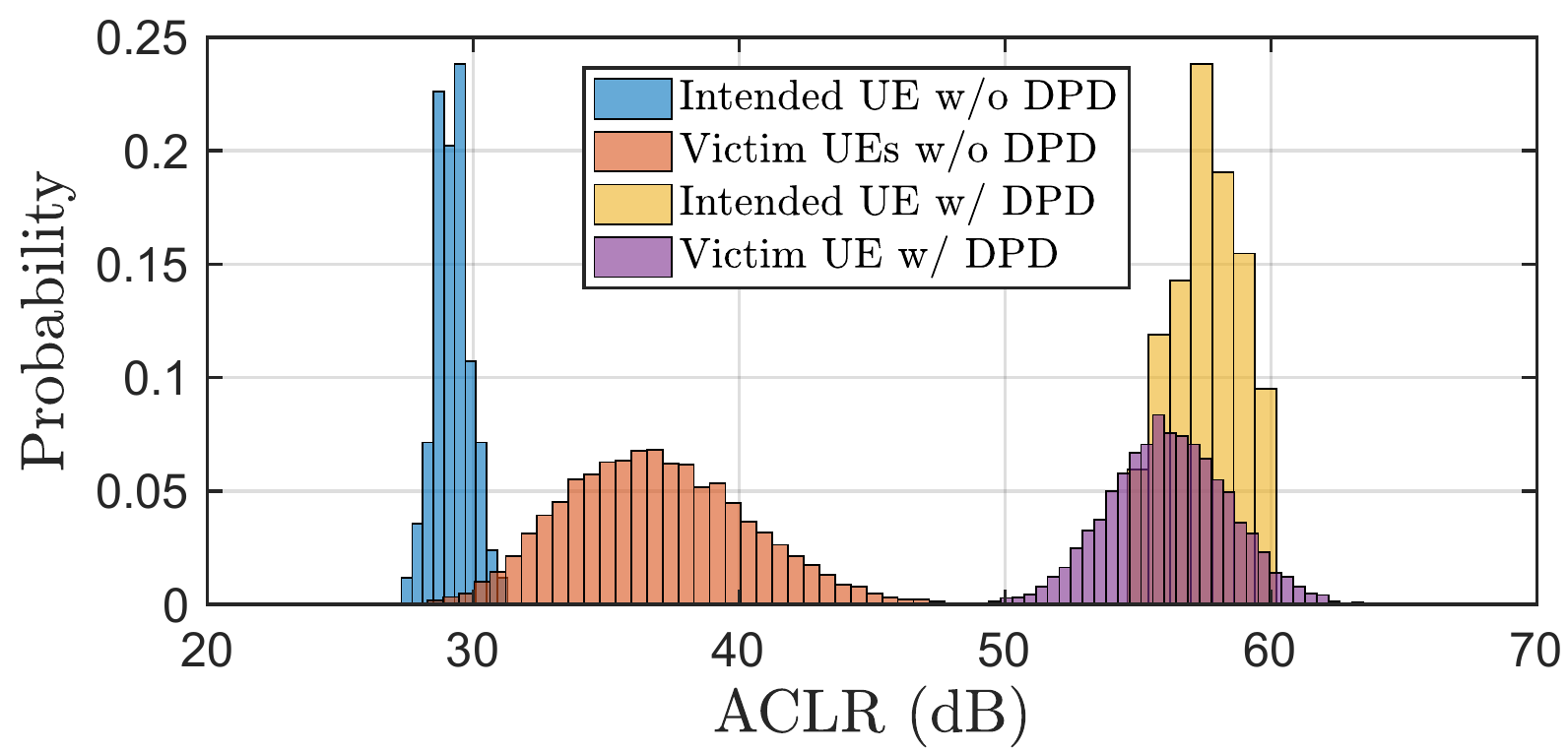}
}
\subfigure{
\includegraphics[width=.75\linewidth]{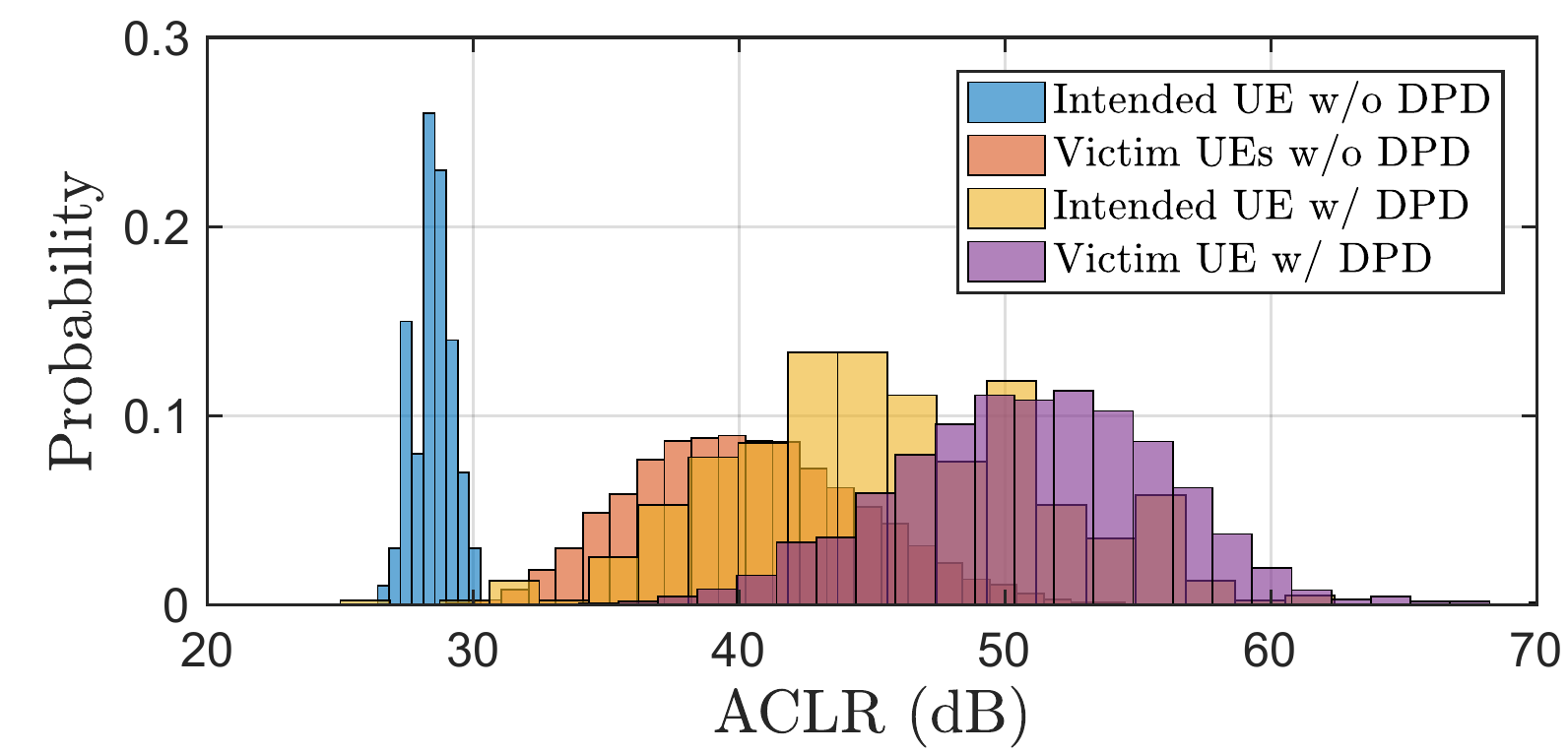}
}
\vspace{-2mm}
\caption{{Empirical ACLR distributions at intended and victim UEs  (\emph{i}) when the DPD parameter learning is carried out separately for each random drop of intended UEs (top) and (\emph{ii}) when the DPD parameter learning is executing only for the first drop while fixed for others (bottom). 'With DPD' refers to the closed-loop DPD method. Baseline PA input crosstalk level is $-$20~dB while the baseline antenna crosstalk level is $-$10~dB.}}
\label{fig:spatial_domain_crosstalk}
\end{figure}
{
\subsection{Results with Crosstalk and Imperfect CSI}\label{sec:imperfect_CSI}
Lastly, we address a yet more practical scenario where imperfect CSI is considered in the analog beamforming and digital precoding stages. For the analog beamforming, we still pursue the multibeam scenario through the optimization approach in \cite{AnalogBeamforming}, but then quantize the resulting phase values with 5-bit resolution to yield the actual beamforming weights. For the digital precoding, in turn, we model the estimation errors and uncertainty in the equivalent beamformed channels at subcarrier level as \cite{ScalingMIMO}
\begin{equation}
    \hat{\mathbf{G}}_{\text{eq}}[k] = \chi{\mathbf{G}}_{\text{eq}}[k]+\sqrt{1-\chi^2}\mathbf{E},
\end{equation}
where $\chi \in [0, 1]$ is a parameter controlling the accuracy of the estimates, while $\mathbf{E}$ is an error matrix whose entries are i.i.d $\mathcal{C}\mathcal{N}(0,1)$ random variables. These estimates are then used to calculate the ZF digital precoder. As a concrete numerical example, the value of $\chi = 0.9$ is used, which corresponds to a power ratio of some $\approx 20  \text{dB}$ between the errors and the true responses. All other parameters, including the crosstalk, are the same as in earlier evaluations.

\begin{figure}[t!]
\centering
\includegraphics[width=.75\linewidth]{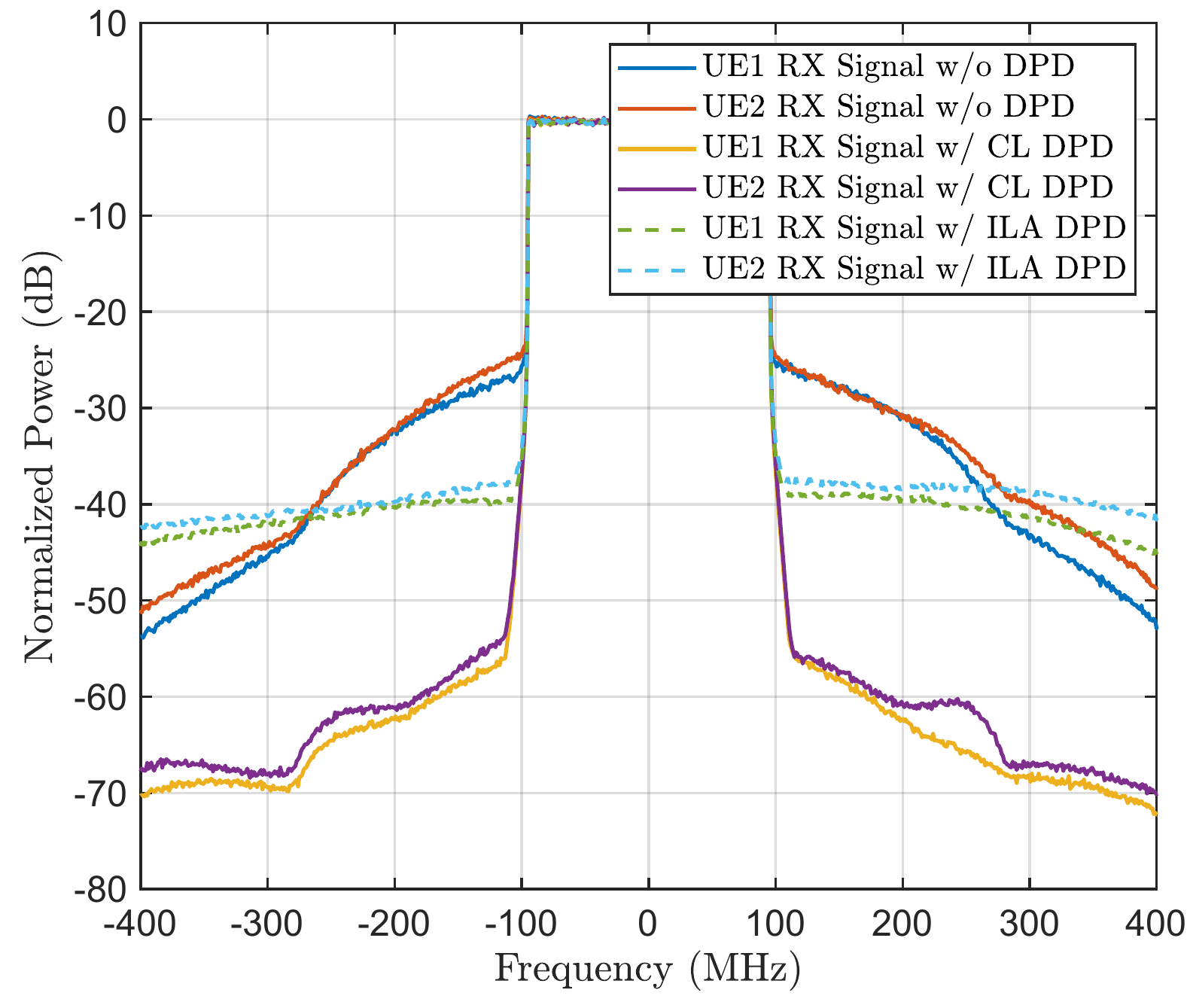}
\caption[]{{Normalized combined spectra at the two intended users, without and with DPD, when the \textit{multi-beam analog beamformers} are adopted. Baseline PA input crosstalk level is $-$20~dB while the baseline antenna crosstalk level is $-$10~dB. Imperfect CSI in terms of quantized analog beamforming and noisy equivalent channel estimates in digital precoding is considered.}}
\label{RX_spectrumsUE_crosstalk_imperfect_csi}
\end{figure}

The obtained results in terms of the example received signal spectra are depicted in Fig. \ref{RX_spectrumsUE_crosstalk_imperfect_csi}. As can be observed, high-quality linearization can still be obtained despite uncertainties in the CSI used for beamforming and precoding. Additionally, the CL DPD is again superior, compared to ILA, primarily due to the crosstalk.}

\section{Conclusions}
\label{sec:Conclusions}
In this article, we addressed the power amplifier (PA) nonlinear distortion problem in future array systems, with specific emphasis on multiuser hybrid beamforming based transmitters at mmWaves. First, assuming the generic case of subcarrier-wise multiuser digital precoding and phase-based single-beam or multi-beam analog beamforming in the involved sub-arrays, together with nonlinear and mutually different PA units with memory, the essential signal models were derived describing the combined or observable nonlinear distortion at receiving ends. Then, stemming from the derived signal models, an efficient single-input DPD architecture and related closed-loop parameter learning solutions were adopted, and shown to allow to simultaneously linearizing the observable signals at all directions where coherent combining takes place. {The nonlinear distortion modeling and corresponding DPD developments were then extended to include also the important practical challenge of mutual coupling, such that PA input crosstalk within each subarray as well as generic antenna coupling within the overall large array were both considered.} 

Specifically, it was shown that the adopted single-input closed-loop DPD unit is capable of suppressing the unwanted emissions stemming from the corresponding subarray towards all the intended receivers, and thus the composite  nonlinear  distortion  observed  at  the  intended  receivers is suppressed by the overall DPD system. {Additionally, it was shown that the system is very robust against crosstalk and that efficient linearization is obtained also from arbitrary victim receivers' point of view, stemming from the combined effect of the DPD system and the array isolation/beamforming.} Extensive numerical performance examples were provided, with specific focus on timely millimeter wave systems, demonstrating and evidencing the excellent linearization performance of the considered approach in different evaluation scenarios {involving measured mmWave PA models. Finally, the DPD system was also shown to be applicable and to offer high-performance linearization when the beamforming stages build on imperfect channel state information.}

\appendices
{\section{Derivation of Received Signal Model (\ref{eq:rx_crosstalk}) under Crosstalk}\label{sec:Appendix}
In this Appendix, the specific analysis steps to derive the received signal model in (\ref{eq:rx_crosstalk}) are provided.First, by expanding the PA output signal expression in (\ref{eq:PA_output_crosstalk}) one gets
\begin{equation}
    \begin{split}
        \bar{y}_{l,m}(n) &= \bar{w}_{l,m}\sum_{\substack{p=1 \\ p, \text{odd}}}^{P}\bar{\alpha}_{l,m,p}(n)\star\psi_{l,p}(n)+ \sum_{\substack{k=1}}^{L}\sum_{\substack{i=1\\i \neq m}}^{M}c_{k,l,m,i}\\
        &\times\Big(w_{k,i} + \sum_{\substack{i=1\\i \neq m}}^{M}b_{k,i,m}w_{k,i}\Big)\sum_{\substack{p=1 \\ p, \text{odd}}}^{P}\bar{\alpha}_{k,i,p}(n)\star\psi_{k,p}(n)
    \end{split}\label{eq:pa_output_appendix}
\end{equation}
Since the coupling factors are generally small, commonly around  $-$10~dB to $-$15~dB \cite{MIMO_DPD_2,MIMO_DPD_3,MIMO_DPD_4,Prediction_distortion,28GHz_crosstalk} or lower, the impact of the terms containing products of the form $c_{k,l,m,i}b_{k,i,m}$ can be assumed to be negligible. Then, the combined received signal at the intended user $u$ reads
\small
\begin{align}
    \begin{split}
    \bar{z}_u(n) &= \sum_{\substack{l=1}}^{L}\sum_{\substack{m=1}}^{M} h_{l,m,u}(n)\star\bar{y}_{l,m}(n)\end{split}\\
    \begin{split}
    &=  \sum_{\substack{l=1}}^{L} h_{l,u}(n)\star\\ &\Big(\sum_{\substack{m=1}}^{M}\Big(e^{j\Delta\beta^{l,l}_{m,m}} + \sum_{\substack{i=1 \\ i \neq m}}^{M} e^{j\Delta\beta^{l,l}_{i,m}}b_{i,m}\Big) \sum_{\substack{p=1 \\ p, \text{odd}}}^{P}\bar{\alpha}_{l,m,p}(n)\star \psi_{l,p}(n)\\ &+\sum_{\substack{m=1}}^{M}\sum_{\substack{k=1}}^{L}\sum_{\substack{i=1\\ i\neq m }}^{M}c_{k,l,i,m} e^{j\Delta{\beta}^{k,l}_{i,m}} \sum_{\substack{p=1 \\ p, \text{odd}}}^{P}\bar{\alpha}_{k,i,p}(n)\star \psi_{k,p}(n)\Big)\label{eq:pre_rx_signal_appendix}
    \end{split}
\end{align}
\normalsize
Then, by combining the signal terms in the last line of (\ref{eq:pre_rx_signal_appendix}) for $k=l$ with those on the second-last line, and summing over $m$, we finally reach
\small
\begin{equation}
 \begin{split}
     \bar{z}_u(n)&=\sum_{\substack{l=1}}^{L} {h^{\text{eff}}_{l,u}(n)}\star\\
     &\Big(\sum_{\substack{p=1 \\ p, \text{odd}}}^{P}\bar{\alpha}^{\text{tot}}_{l,p}(n)\star\psi_{l,p}(n) +   \sum_{\substack{k=1 \\ k\neq l }}^{L}\sum_{\substack{p=1 \\ p, \text{odd}}}^{P}{f}^{\text{tot}}_{l,k,p}(n)\star\psi_{k,p}(n)\Big)\label{RX_signal1_appendix},
 \end{split}
\end{equation}
\normalsize
where $\bar{\alpha}^{\text{tot}}_{l,p}(n)$ and ${f}^{\text{tot}}_{l,k,p}(n)$ are as defined below (\ref{eq:rx_crosstalk}). This concludes the proof.}

\bibliographystyle{IEEEbib}
\bibliography{Ref}
\begin{IEEEbiography}[{\includegraphics[width=1in,height=1.25in,clip,keepaspectratio]{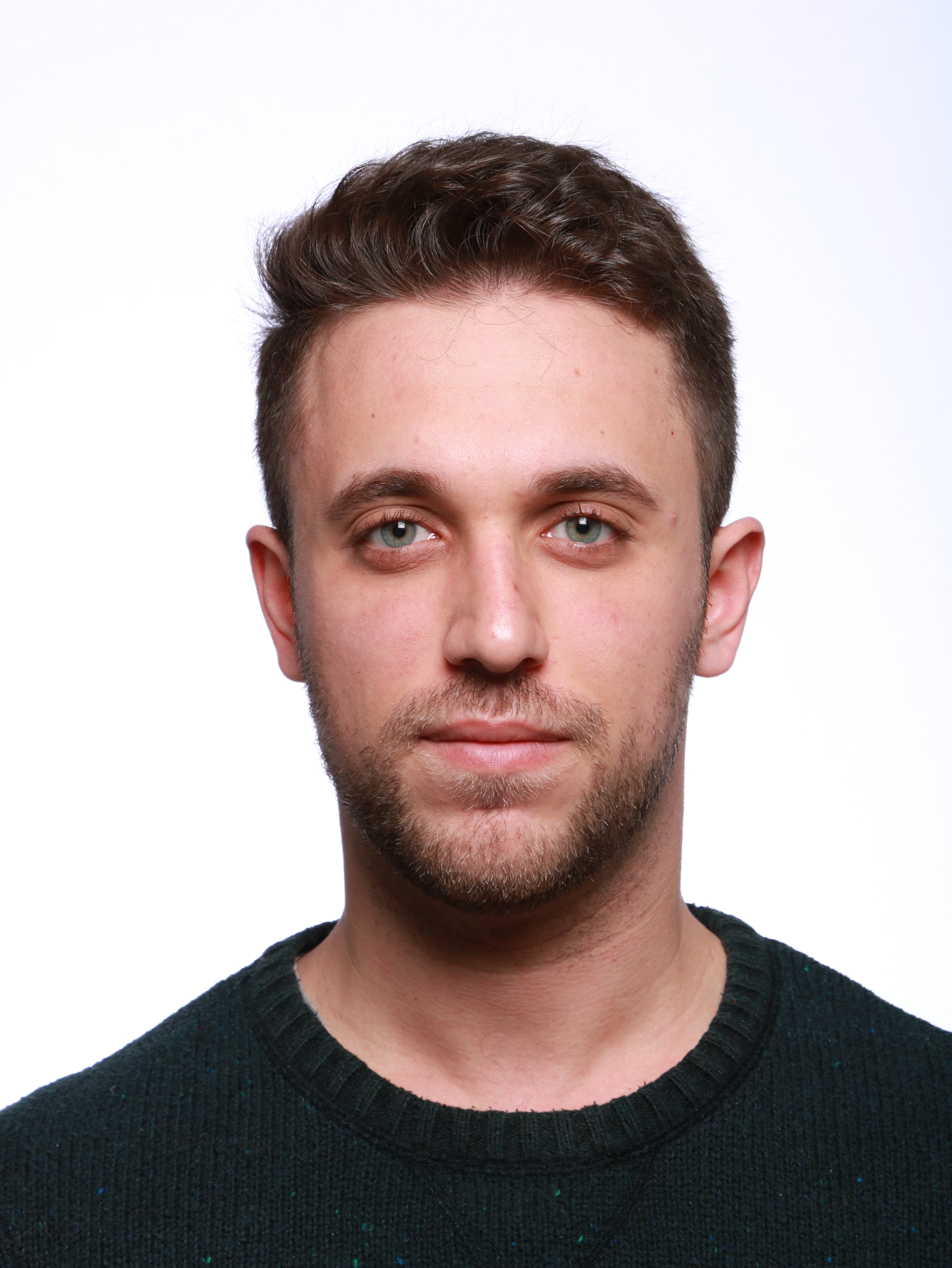}}]{Alberto Brihuega} (S'18) received the B.Sc. and M.Sc. degrees in Telecommunications Engineering from Universidad Polit\'ecnica de Madrid, Spain, in 2015 and 2017, respectively. He is currently working towards the Ph.D. degree with Tampere University, Finland, where he is a researcher with the Department of Electrical Engineering. His research interests include statistical and adaptive digital signal processing for compensation of hardware impairments in large-array antenna transceivers.
\end{IEEEbiography}

\begin{IEEEbiography}[{\includegraphics[width=1in,height=1.25in,clip,keepaspectratio]{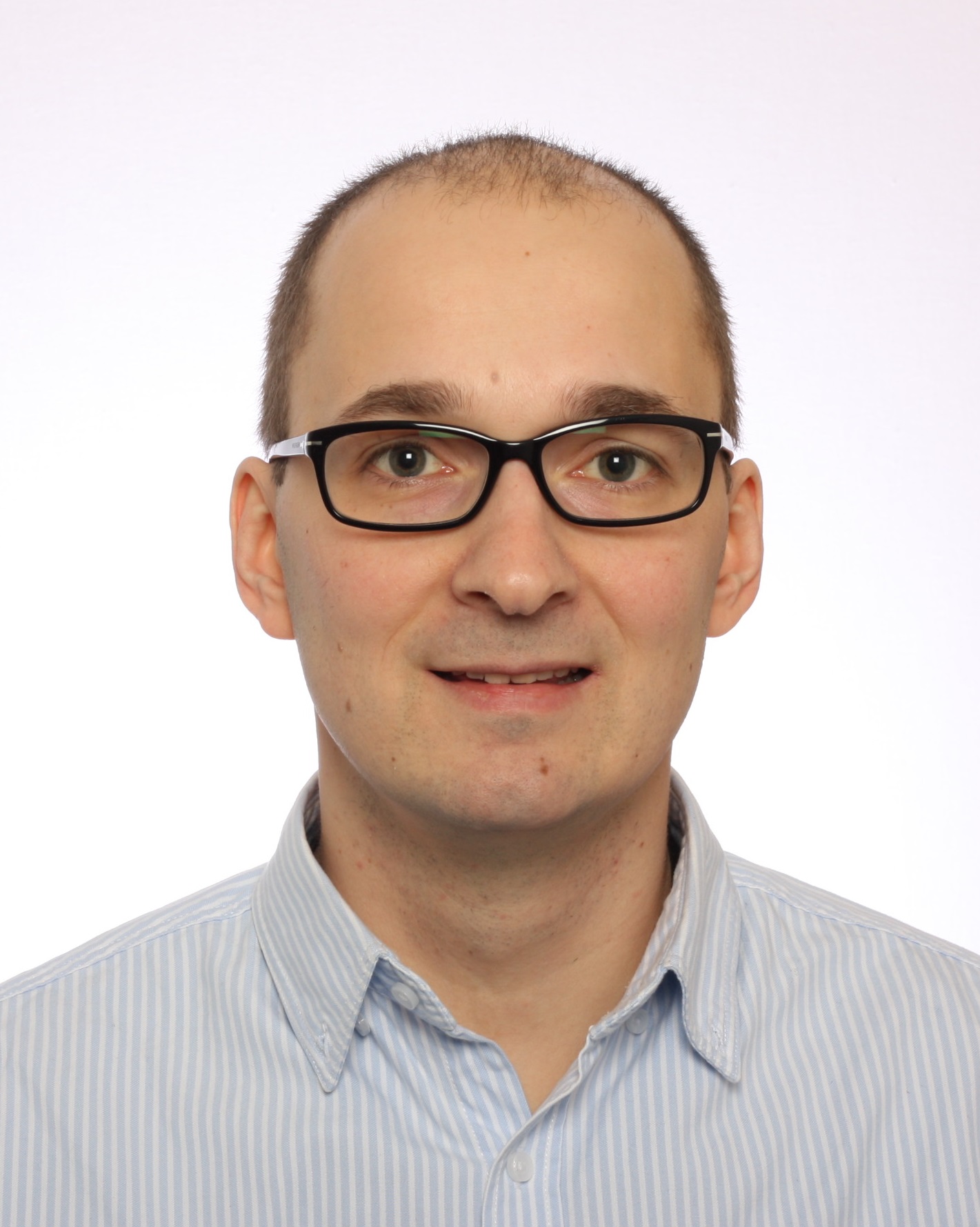}}]{Lauri Anttila} received the M.Sc.and D.Sc. (Hons.) degrees in electrical engineering from Tampere University of Technology (TUT), Tampere, Finland, in 2004 and 2011, respectively. Since 2016, he has been a University Researcher with the Department of Electrical Engineering, Tampere University (formerly TUT). From 2016 to 2017, he was a Visiting Research Fellow with the Department of Electronics and Nanoengineering, Aalto University, Helsinki, Finland. He has coauthored over 100 refereed articles and 3 book chapters. His current research interests include radio communications and signal processing, with a focus on the radio implementation challenges in systems such as 5G, full-duplex radio, and large-scale antenna systems.
\end{IEEEbiography}

\begin{IEEEbiography}[{\includegraphics[width=1in,height=1.25in,clip,keepaspectratio]{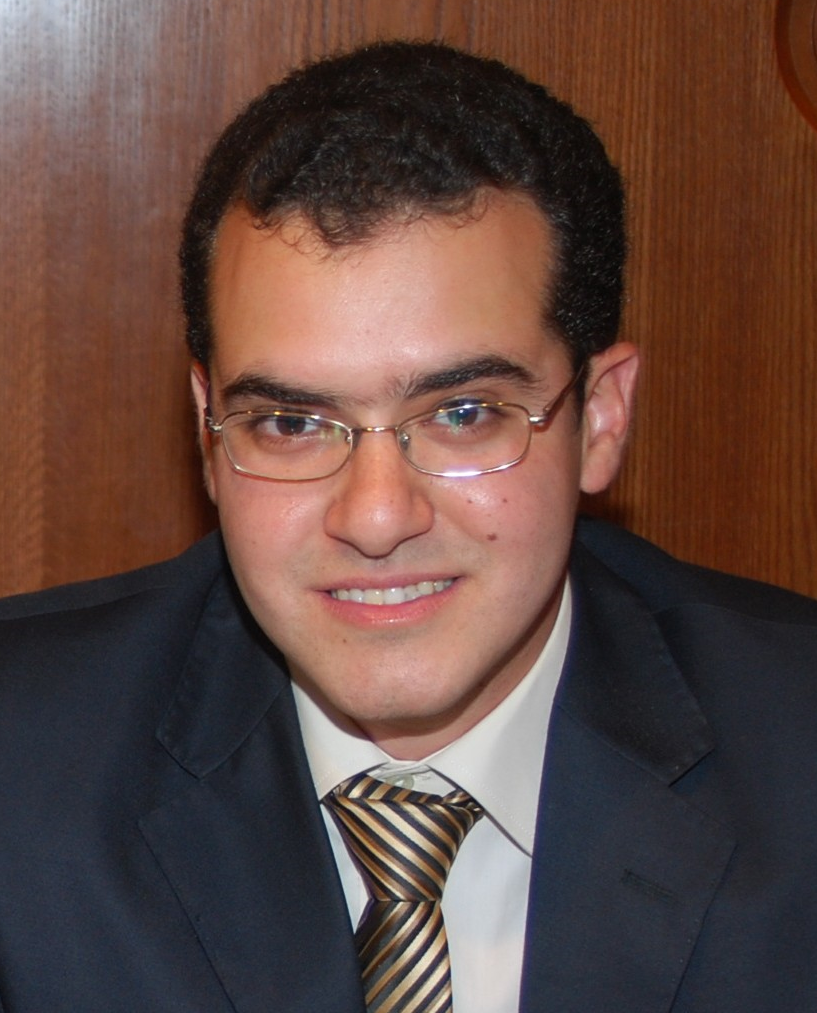}}]{Mahmoud Abdelaziz} received the D.Sc. degree (with honors) in Electronics and Communications Engineering from Tampere University of Technology, Finland, in 2017. He then worked as a Postdoctoral Researcher at the same University. He is currently an Assistant Professor at Zewail City of Science and Technology.  From 2007 to 2012 he has worked as a communications and signal processing engineer as well as an embedded systems engineer at Newport Media Inc., Etisalat Egypt, and Axxcelera Broadband Wireless. His research interests include statistical and adaptive signal processing in flexible radio transceivers, in particular, behavioral modelling and digital pre-distortion of power amplifiers in single and multiple antenna transmitters.
\end{IEEEbiography}

\begin{IEEEbiography}[{\includegraphics[width=1in,height=1.25in,clip,keepaspectratio]{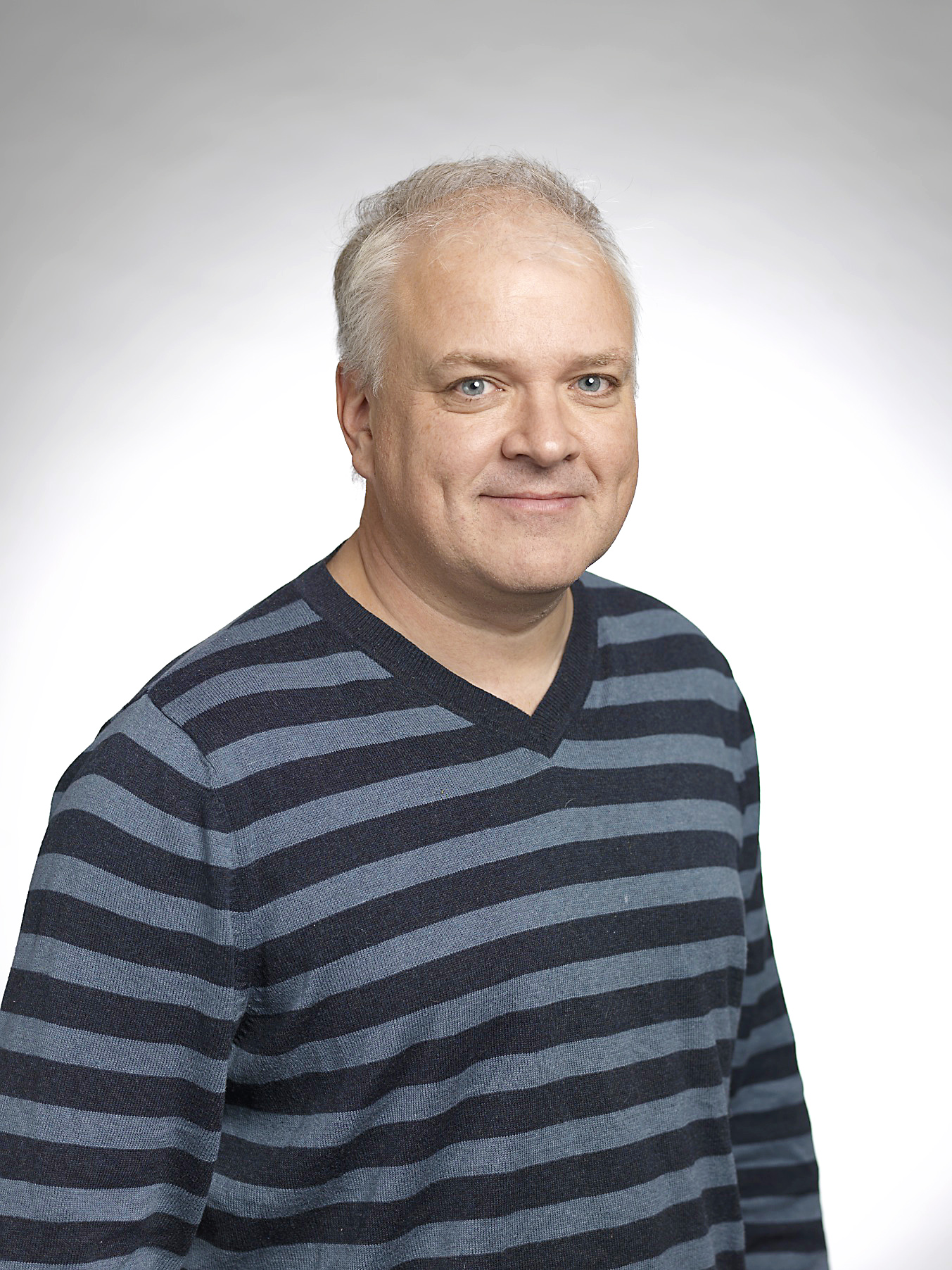}}]{Thomas Eriksson} received the Ph.D. degree in Information Theory from Chalmers University of Technology, Gothenburg, Sweden, in 1996. From 1990 to 1996, he was at Chalmers. In 1997 and 1998, he was at AT\&T Labs - Research, Murray Hill, NJ, USA. In 1998 and 1999, he was at Ericsson Radio Systems AB, Kista, Sweden. Since 1999, he has been with Chalmers University, where he is currently a professor of communication systems. Further, he was a guest professor with Yonsei University, S. Korea, in 2003-2004. He has authored or co-authored more than 250 journal and conference papers, and holds 14 patents. Prof. Eriksson is leading the research on hardware-constrained communications with Chalmers University of Technology. His research interests include communication, data compression, and modeling and compensation of non-ideal hardware components (e.g. amplifiers, oscillators, and modulators in communication transmitters and receivers, including massive MIMO). Currently, he is leading several projects on e.g. 1) massive MIMO communications with imperfect hardware, 2) MIMO communication taken to its limits: 100Gbit/s link demonstration, 3) wideband linearization (DPD), 4), Satellite communication with phase noise limitations, 5) Efficient and linear transceivers, etc. He is currently the Vice Head of the Department of Signals and Systems with Chalmers University of Technology, where he is responsible for undergraduate and masters education.
\end{IEEEbiography}

\begin{IEEEbiography}[{\includegraphics[width=1in,height=1.25in,clip,keepaspectratio]{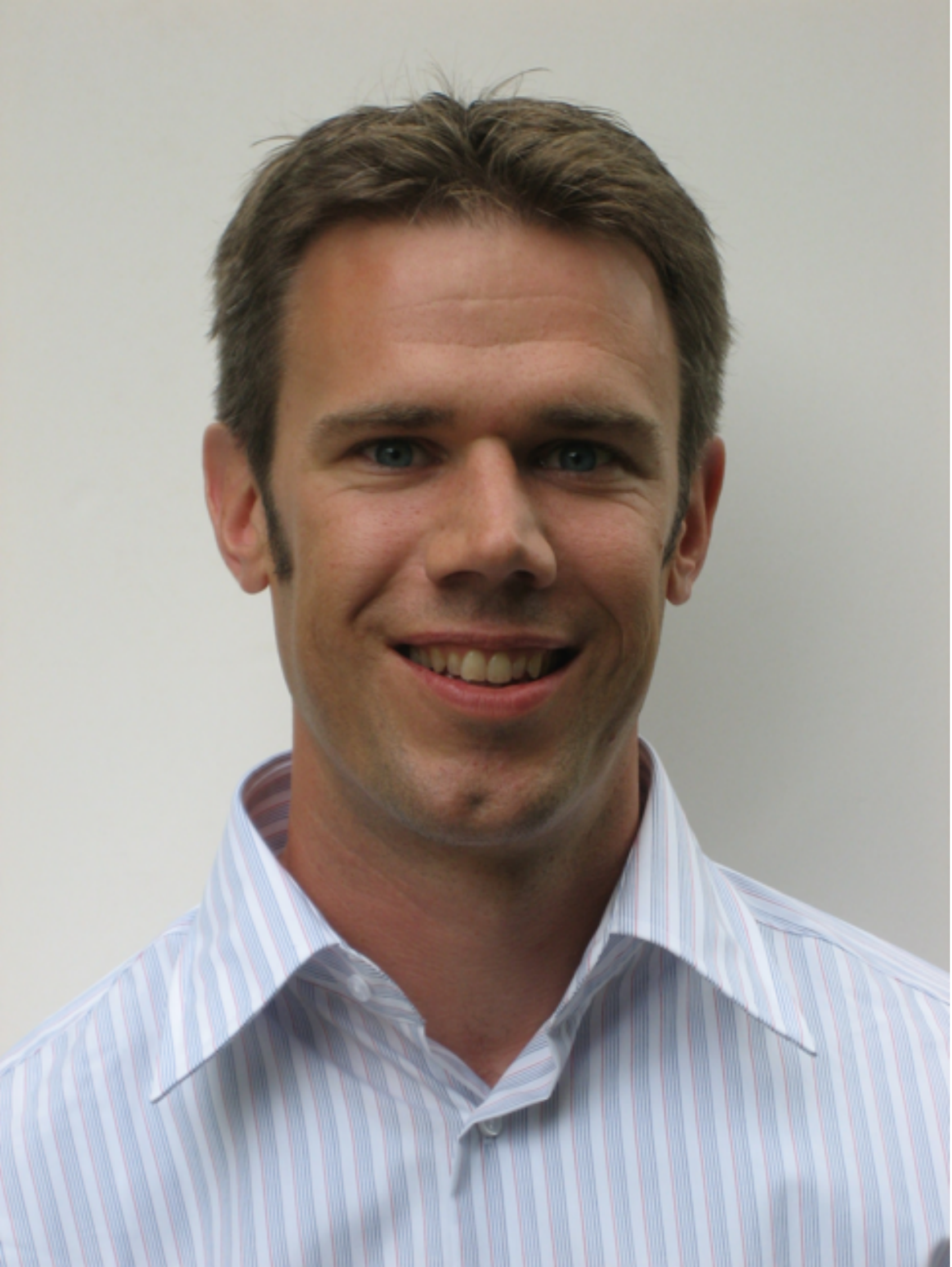}}]{Fredrik Tufvesson} received his Ph.D. in 2000 from Lund University in Sweden. After two years at a startup company, he joined the department of Electrical and Information Technology at Lund University, where he is now professor of radio systems. His main research interests is the interplay between the radio channel and the rest of the communication system with various applications in 5G systems such as massive MIMO, mm wave communication, vehicular communication and radio based positioning.
\end{IEEEbiography}

\begin{IEEEbiography}[{\includegraphics[width=1in,height=1.25in,clip,keepaspectratio]{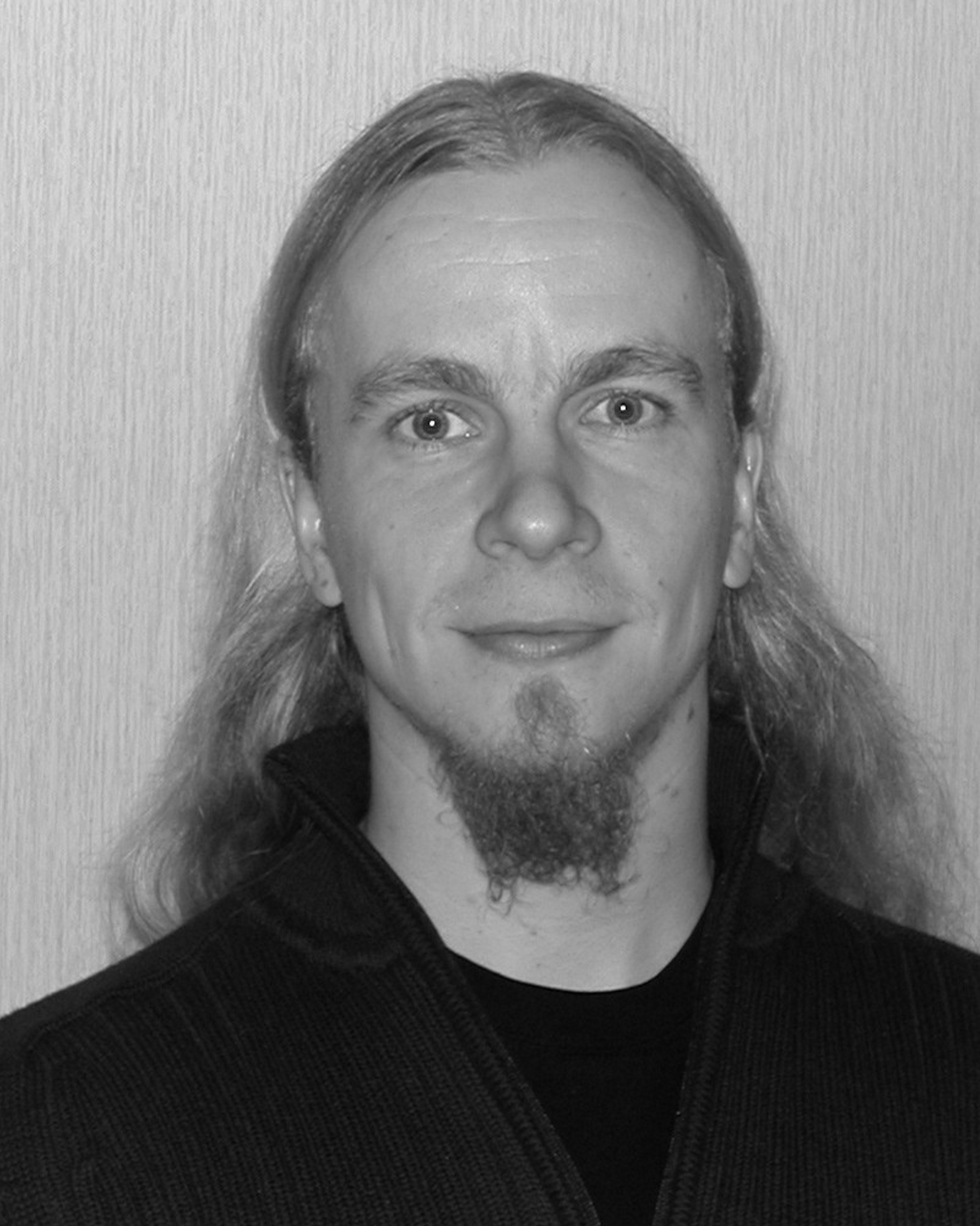}}]{Mikko Valkama} (S'00--M'01--SM'15) received the M.Sc.(Tech.) and D.Sc.(Tech.) degrees in electrical engineering (EE) from the Tampere University of Technology (TUT), Tampere, Finland, in 2000 and 2001, respectively. In 2003, he was a Visiting Post-Doctoral Research Fellow with the Communications Systems and Signal Processing Institute, San Diego State University (SDSU), San Diego, CA, USA. He is currently a Full Professor and the Department Head of Electrical Engineering with the newly established Tampere University (TAU), Tampere, Finland. His current research interests include radio communications, radio localization, and radio-based sensing, with particular emphasis on 5G and beyond mobile radio networks. Dr. Valkama was a recipient of the Best Ph.D. Thesis Award of the Finnish Academy of Science and Letters for his Ph.D. dissertation.
\end{IEEEbiography}
 
\end{document}